\documentclass[12pt,preprint]{article}
\usepackage{amssymb}
\usepackage{amsmath}
\usepackage{graphics}
\usepackage{epsfig}
\usepackage{psfig}

\newcommand{\eqb}{\begin{equation}}
\newcommand{\eqe}{\end{equation}}
\newcommand{\dmb}{\begin{displaymath}}
\newcommand{\dme}{\end{displaymath}}

\newcommand{\eab}{\begin{eqnarray}}
\newcommand{\eae}{\end{eqnarray}}

\newcommand{\be}{\begin{equation}}
\newcommand{\ee}{\end{equation}}

\begin{document}

\begin{titlepage}
\begin{flushright} 
\end{flushright}
\vspace{0.6cm}

\begin{center}
\Large{Strong and electromagnetic decays of 
the light scalar mesons interpreted as tetraquark states}
\vspace{1.5cm}

\large{Francesco Giacosa}

\end{center}
\vspace{1.5cm} 

\begin{center}
{\em Institut f\"ur Theoretische Physik\\ 
Universit\"at Frankfurt\\ 
Johann Wolfgang Goethe - Universit\"at\\ 
Max von Laue--Str. 1\\ 
60438 Frankfurt, Germany}
\end{center}
\vspace{1.5cm}

\begin{abstract}
The study of two-pseudoscalar and two-photon decays for the scalar meson nonet below 1 GeV 
is performed within an effective approach in which the scalar resonances
are described as (Jaffe's) tetraquark states. 
The dominant (fall apart decay) and the subdominant 
(one transverse gluon as intermediate state) decay amplitudes are systematically taken 
into account.
The latter improves the agreement with the experimental data. 
Possible scenarios concerning  the scalar-isoscalar mixing are discussed.
\end{abstract} 

\end{titlepage}

\bigskip

\section{Introduction}

The interpretation of the mesonic scalar states below 2 GeV is not yet
univocal established \cite{amslerrev,closerev,mink}. According to the most
popular scenario, one interprets the isovector and isotriplet resonances $%
a_{0}(1400)$ and $K(1430)$ as the ground-state quark-antiquark bound states.
The three isoscalar resonances $f_{0}(1300)$, $f_{0}(1500)$ and $f_{0}(1710)$
are a mixture of two isoscalar quarkonia and bare glueball configurations
(we refer to \cite%
{amslerrev,closerev,close95,weing,gutsche,closekirk,giacosa,giacosachiral}
and Refs. therein). As a consequence, the scalar states below 1 GeV ($\sigma
,$ $k,$ $f_{0}(980)$ and $a_{0}(980)$) must be something else, like (loosely
bound) mesonic molecular states \cite{mesonicmol1,mesonicmol2} or Jaffe's
tetraquark states \cite{amslerrev,jaffe,exotica,maiani}. In this work we
explore by means of an effective approach the phenomenological implications
of Jaffe's states, whose building blocks are a diquark ($q^{2}$) and an
antidiquark ($\overline{q}^{2}$); calculations based on one-gluon exchange
support a strong attraction among two quarks in a color antitriplet ($%
\overline{3}_{C}$), a flavor antitriplet ($\overline{3}_{F}$) and spinless
configuration \cite{amslerrev,jaffe} (color and flavor triplets are realized
for an antidiquark). Naively speaking, a diquark `behaves like an antiquark'
from a flavor (and color) point of view \cite{exotica}:%
\begin{equation}
\lbrack u,d]\leftrightarrow \overline{s},\text{ }[u,s]\leftrightarrow 
\overline{d},\text{ }[d,s]\leftrightarrow \overline{u},
\end{equation}%
therefore out of a diquark and an antidiquark one can build a full scalar
nonet, whose most appealing property is the reversed mass-order, thus
explaining the (almost) degeneracy of the isoscalar state $f_{0}(980)$
(whose dominant contribution is the tetraquark structure \textquotedblleft $%
\overline{s}s(\overline{u}u+\overline{d}d)$\textquotedblright ) and the
isovector $a_{0}(980)$ (the neutral one interpreted as \textquotedblleft $%
\overline{s}s(\overline{u}u-\overline{d}d)$\textquotedblright ); the $\sigma 
$ (dominantly \textquotedblleft $\overline{u}u\overline{d}d$%
\textquotedblright ) is then the lightest, in between one expects the kaonic
state $k(800)$ ($k^{+}$ interpreted as \textquotedblleft $\overline{d}d%
\overline{s}u$\textquotedblright ), which is omitted in the compilation of
PDG \cite{pdg}, but listed in many recent theoretical and experimental works
(\cite{vanbeveren,ishida,black,aitala,buggexp05,bugg,buggexp06,ablikim} and
Refs. therein).

Evidence for a broad $\sigma $ state is also found, together with $%
a_{0}(980) $ and $f_{0}(980),$ in the theoretical work of \cite%
{dobado,olleroset} where a unitarized ChPT is used; as shown in \cite%
{ollerosetpelaez,pelaez} a broad scalar state $k$ exists as well. The fact
that a full scalar nonet is generated within the same approach points to a
similar inner structure of the low-lying scalar states. Furthermore, the
large-$N_{c}$ behavior of the light scalar states indicates large non-$%
\overline{q}q$ amounts in their spectroscopic wave function \cite%
{pelaezprl,pelaez}. Support for the Jaffe's states has also been found in
Lattice calculations \cite{jaffelatt,okiharu}, where the diquark $q^{2}$ and
the antidiquark $\overline{q}^{2}$ are connected by a flux tube.

In \cite{fariborz,napsuciale} the scalar states below $1$ GeV have also a
dominant tetraquark structure; mixing among light tetraquark states and
heavy quarkonia states, generating two scalar nonet of mixed states below
and above $1$ GeV is described.

In the present work we intend to analyze the two-pseudoscalar decays of the
scalar states below 1 GeV when interpreted as Jaffe's tetraquark states; to
this end we consider the dominant and the subdominant diagrams in the large-$%
N_{c}$ expansion which describe the transition of a tetraquark scalar state
into two pseudoscalars \cite{jaffe,maiani}.

The dominant diagram, depicted in Fig. 1.a, occurs by switch of a quark
belonging to the compact diquark with an antiquark of the anti-diquark, thus
generating two $\overline{q}q$ objects, which separate as pseudoscalar
mesons: $q^{2}\overline{q}^{2}\rightarrow (q\overline{q})(q\overline{q}%
)\rightarrow (q\overline{q})+(q\overline{q})$. The two pseudoscalar mesons
fall apart from the tetraquark configuration; this decay mechanism was
denoted as OZI-superallowed in Ref. \cite{jaffe}.

The subdominant diagram, depicted in Fig. 1.b, occurs via an annihilation of
a quark and an antiquark (into one gluon), with subsequent $\overline{q}q$
creation and two-pseudoscalar decay: $q^{2}\overline{q}^{2}\rightarrow qg%
\overline{q}\rightarrow (q\overline{q})+(q\overline{q})$. Although
suppressed by a factor $N_{c}$, the fact that the annihilation of the
quark-antiquark pair can occur with only one gluon as intermediate state (as
already noted in Ref. \cite{jaffe}), i.e. to order $\alpha _{s}$ only, may
indeed indicate that the corresponding amplitudes are not negligible.

In Ref. \cite{jaffe} the subdominant coupling has not been considered in the
decay rates; in \cite{maiani} it has been introduced as the last step of the
analysis in order to improve the results of the superallowed decays. In the
present work we intend to systematically write down the expressions for all
scalar-to-pseudoscalars transition amplitudes as functions of the strengths
of the dominant ("fall apart") and subdominant ("one intermediate gluon")
diagrams, denoted as $c_{1}$ and $c_{2}$ respectively. Then, by a fitting
procedure to experimental known branching ratios, we determine the quantity $%
c_{2}/c_{1},$ which measures the intensity of the subdominant decay
mechanism with respect to the dominant one. Two possible solutions with a
non-zero value of the ratio $c_{2}/c_{1}$ are discussed, which in turn
correspond to different isoscalar mixing configurations.

An interesting feature of the subdominant decay is the strong enhancement of
the coupling $g_{f_{0}\rightarrow \overline{K}K}^{2}$ with respect to $%
g_{a_{0}\rightarrow \overline{K}K}^{2},$ thus possibly solving the problem
of Jaffe's model mentioned in \cite{bugg} at the leading (OZI-superallowed,
Fig. 1.a) order for which $g_{f_{0}\rightarrow \overline{K}%
K}^{2}/g_{a_{0}\rightarrow \overline{K}K}^{2}=1$ (if no isoscalar mixing
occurs, even smaller if mixing is introduced), in clear contrast with the
result $g_{f_{0}\rightarrow \overline{K}K}^{2}/g_{a_{0}\rightarrow \overline{%
K}K}^{2}=2.15\pm 0.40$ reported in the analysis of Refs. \cite%
{bugg,buggexp06}. The subdominant decay mechanism of Fig. 1.b can explain
the experimental value without introducing explicit $\overline{K}K$ clouds
dressing the scalar resonances.

We then turn our attention to the decay of scalar states into two photons,
which in line of the previous discussion is described by two decay
mechanisms; they are analogous to the strong decay diagrams of Fig. 1, where
one replaces the two pseudoscalar mesons in the final state with two
photons. We present the theoretical ratios and the phenomenological
discussion.

The paper is organized as follows: in section 2 we write down the Lagrangian
for the description of the scalar mesons below $1$ GeV as tetraquark states;
the mass term and the two-body strong decays are presented. In section 3 we
perform a phenomenological study with the available experimental data. In
section 4 we describe the two-photon transitions and in section 5 we drive
our conclusions.

\section{The model}

\subsection{The Lagrangian}

The basic terms are the pseudoscalar fields, collected in the matrix $%
\mathcal{P}=\frac{1}{\sqrt{2}}\sum_{i=0}^{8}P^{i}\lambda _{i}\,$(the $%
\lambda _{i}$ are Gell-Mann matrices), and low-lying scalar fields,
collected in the matrix $\mathcal{S}$ defined as (see Appendix A):%
\begin{eqnarray}
\mathcal{S} &\mathcal{=}&\left( 
\begin{array}{ccc}
\sigma _{B} & k^{0} & k^{+} \\ 
\overline{k}^{0} & \sqrt{\frac{1}{2}}(f_{B}+a_{0}^{0}) & a_{0}^{+} \\ 
k^{-} & a_{0}^{-} & \sqrt{\frac{1}{2}}(f_{B}-a_{0}^{0})%
\end{array}%
\right)  \notag \\
&=&\left( 
\begin{array}{ccc}
\frac{1}{2}[u,d][\overline{u},\overline{d}] & \frac{1}{2}[u,d][\overline{u},%
\overline{s}] & \frac{1}{2}[u,d][\overline{d},\overline{s}] \\ 
\frac{1}{2}[u,s][\overline{u},\overline{d}] & \frac{1}{2}[u,s][\overline{u},%
\overline{s}] & \frac{1}{2}[u,s][\overline{d},\overline{s}] \\ 
\frac{1}{2}[d,s][\overline{u},\overline{d}] & \frac{1}{2}[d,s][\overline{u},%
\overline{s}] & \frac{1}{2}[d,s][\overline{d},\overline{s}]%
\end{array}%
\right) ,  \label{S}
\end{eqnarray}%
where in the second matrix the diquark-antidiquark decomposition has been
made explicit. The states $\sigma _{B}=\frac{1}{2}[u,d][\overline{u},%
\overline{d}]$ and $f_{B}=\frac{1}{2\sqrt{2}}([u,s][\overline{u},\overline{s}%
]+[d,s][\overline{d},\overline{s}])$ refer to $bare$ (unmixed) states. A
mixing of these configurations, leading to the physical states $\sigma $ and 
$f_{0}(980)$, is possible and considered below.

The Lagrangian for the scalar-pseudoscalar interaction reads:%
\begin{eqnarray}
\mathcal{L} &\mathcal{=}&\left\langle \frac{1}{2}\mathcal{(\partial }_{\mu }%
\mathcal{P)}^{2}-\mathcal{P}^{2}\chi _{P}\right\rangle +\mathcal{L}%
_{mix}^{P}+\left\langle \frac{1}{2}\mathcal{(\partial }_{\mu }\mathcal{S)}%
^{2}-\mathcal{S}^{2}\chi _{S}\right\rangle \mathcal{+L}_{mix}^{S}  \notag \\
&&+c_{1}\mathcal{S}_{ij}\left\langle A^{i}\mathcal{P}^{t}A^{j}\mathcal{P}%
\right\rangle -c_{2}\mathcal{S}_{ij}\left\langle A^{i}A^{j}\mathcal{P}%
^{2}\right\rangle ,  \label{lag}
\end{eqnarray}%
where $\left\langle ...\right\rangle $ denotes trace over flavor. Some
comments are in order:

$\bullet $ In the first term the quantity $\chi _{P}=B\cdot
diag\{m_{u},m_{d}=m_{u},m_{s}\}$ encodes flavor symmetry violation. It
corresponds to the lowest order chiral perturbation theory result \cite{chpt}
(see also \cite{scherer} and Refs. therein), to which we refer for a careful
description. The second term $\mathcal{L}_{mix}^{P}=-\frac{\gamma _{P}}{2}%
\left( P^{0}\right) ^{2}-z_{P}P^{0}P^{8}$ takes into account the enhanced
flavor-singlet mass ($U_{A}(1)$ anomaly) and the octet-singlet mixing,
leading to the physical states (we follow the notations of \cite%
{giacosachiral})%
\begin{equation}
\eta \,=\,P^{8}\cos \theta _{P}\,-\,P^{0}\sin \theta _{P}\,,\hspace*{0.25cm}%
\eta ^{\prime }\,=P^{8}\,\sin \theta _{P}\,\ +\,P^{0}\,\cos \theta _{P},
\label{etas}
\end{equation}%
where $\theta _{P}$ is the pseudoscalar mixing angle. According to the the
standard procedure~\cite{ecker,cirigliano,venugopal} we diagonalize the
corresponding $\eta ^{0}$-$\eta ^{8}$ mass matrix to obtain the masses of $%
\eta $ and $\eta ^{\prime }$. By using $M_{\pi }=139.57$~MeV, $M_{K}=493.677$%
~MeV (the physical charged pion and kaon masses), $M_{\eta }=547.75$~MeV and 
$M_{\eta ^{\prime }}=957.78$~MeV the mixing angle is determined as $\theta
_{P}=-9.95^{\circ }$, which corresponds to the tree-level result (see
details in Ref.~\cite{venugopal}). Correspondingly one finds $M_{P^{0}}=%
\sqrt{\left( M_{\pi }^{2}+2M_{K}^{2}\right) /3+\gamma _{P}}=948.10$ MeV and $%
z_{P}=-0.105$ GeV $^{2}$ \cite{giacosachiral}.

$\bullet $ The third and the fourth terms of Eq. (\ref{lag}) refer to the
quadratic part of the scalar tetraquark nonet; $\chi _{S}=diag\{\alpha
,\beta ,\beta \}$ and $\alpha <\beta $ takes into account that the diquark $%
[u,d]$ is lighter than the partners (due to $m_{u}=m_{d}<m_{s}$). In this
way the masses are given as 
\begin{equation}
M_{a_{0}}^{2}=M_{f_{B}}^{2}=2\beta ,\text{ }M_{k}^{2}=(\alpha +\beta ),\text{
}M_{\sigma _{B}}^{2}=2\alpha ,  \label{scalmass}
\end{equation}%
corresponding to the reversed mass ordering. The fourth term $\mathcal{L}%
_{mix}^{S}$describes the mixing of $\sigma _{B}$ and $f_{B}$; details about
it are presented in the next subsection.

It is interesting to note that the same (inverted) mass ordering for the
light scalars can be obtained in the framework of the $SU(3)$ linear $\sigma 
$ model \cite{juergen} with $U_{A}(1)$-anomaly.

It is important to stress that in this work we consider only tree-level
expression for the decays (see Sections 2.3 and 3) and no mass-shift via
loop diagrams is calculated; the Lagrangian-masses for the state $a_{0}(980)$
and $k$ reported in Eq. (\ref{scalmass}) (derived from the Lagrangian (\ref%
{lag})) are therefore the physical masses in our phenomenological approach.
Similarly, $M_{f_{B}}$ and $M_{\sigma _{B}}$ of Eq. (\ref{scalmass}) would
be physical masses if no isoscalar mixing among $\sigma _{B}$ and $f_{B}$
would occur. The physical masses for the states $\sigma $ and $f_{0}(980)$
are determined in the next subsection when isoscalar mixing is considered.

$\bullet $ The last two terms of Eq. (\ref{lag}) correspond to the large-$%
N_{c}$ dominant (proportional to $c_{1}$) and subdominant (proportional to $%
c_{2}$) two-pseudoscalar decays of the low-lying scalars when interpreted as
tetraquark states. The matrices $A^{i}$ entering in the expression (\ref{lag}%
) are the three antisymmetric real $3\times 3$ matrices:%
\begin{equation}
A^{1}=i\lambda _{2},\text{ }A^{2}=i\lambda _{5},\text{ }A^{3}=i\lambda _{7}.
\label{A}
\end{equation}

Both terms are $SU(3)$-flavour invariant; the details about the
transformation properties are reported in Appendix A, where also the other
flavour invariant terms are listed (Eq. (\ref{all})).

We do not consider in the present work decay terms which break flavour
symmetry in virtue of $\alpha \neq \beta $ (Eq. (\ref{scalmass})). Generally
flavour-breaking corrections to the decay amplitudes are not large and do
not change the qualitative picture; furthermore the consideration of such
terms would imply a too large number of parameters for the three-level decay
amplitudes. For these reasons the consideration of such terms is beyond the
goal of the present paper, but it represents a possible future development
of our study.

$\bullet $ As described in Appendix A, the $S\rightarrow PP$ interaction
term $c_{3}\mathcal{S}_{ij}\left\langle A^{i}A^{j}\mathcal{P}\right\rangle
\left\langle \mathcal{P}\right\rangle $ is also suppressed of a factor $%
N_{c} $ with respect to the dominant OZI-superallowed one. It is coupled to
the flavour-blind pseudoscalar configuration $\left\langle \mathcal{P}%
\right\rangle =\sqrt{3}P^{0}$, connected to the physical fields $\eta $ and $%
\eta ^{\prime }$ in equation (\ref{etas}). It can however occur with at
least two intermediate transverse gluons attached to $P^{0}$, i.e. to order $%
\alpha _{s}^{2}.$ It's contribution is then believed to be smaller than the
diagram of Fig. 1.b, which takes place at order $\alpha _{s}.$ A gluonic
amount in the wave functions of the $\eta $ and $\eta ^{\prime }$ states
would enhance this channel, but such an eventuality seems not to occur \cite%
{klempt,Benayoun:1999au,bini}. Furthermore, in the analysis of \cite{tensor}%
, where the two-pseudoscalar decays of the experimentally well-known tensor
mesons are evaluated, a good description of data is obtained without an
enhanced flavor-blind channel in the pseudoscalar mesonic sector. These
arguments lead us not to consider the term $c_{3}\mathcal{S}%
_{ij}\left\langle A^{i}A^{j}\mathcal{P}\right\rangle \left\langle \mathcal{P}%
\right\rangle $ in the present work. According to our view, the further
systematical inclusion of this term, which affects the couplings involving
the $\eta $ and $\eta ^{\prime }$ mesons only, would be then necessary when
the experimental knowledge on the light scalar meson sector becomes more
exhaustive.

\subsection{Mixing}

We discuss the term $\mathcal{L}_{mix}^{S}$ of Eq. (\ref{lag}) which
generates mixing between $\sigma _{B}$ and $f_{B}.$ In line with the
pseudoscalar sector we consider the flavor singlet and octet four-quark
configurations $S_{0}=\sqrt{2/3}f_{B}+\sqrt{1/3}\sigma _{B}$ and $S_{8}=%
\sqrt{1/3}f_{B}-\sqrt{2/3}\sigma _{B}$ and define 
\begin{equation}
\mathcal{L}_{mix}^{S}=-\frac{1}{2}\gamma _{S}S_{0}^{2}-z_{S}S_{0}S_{8},
\label{lsmix}
\end{equation}%
where a octect-singlet mixing and mass-modification for the flavor blind
state are taken into account; we also refer to \cite{maiani} for this point.

By using Eqs. (\ref{lag}) and (\ref{lsmix}) the quadratic part referring to
the isoscalar states $f_{B}$ and $\sigma _{B}$ and their relative mixing
reads: 
\begin{equation}
\mathcal{L}_{isoscalar}^{S}=-\frac{1}{2}\left( 
\begin{array}{cc}
\sigma _{B} & f_{B}%
\end{array}%
\right) \Omega \left( 
\begin{array}{c}
\sigma _{B} \\ 
f_{B}%
\end{array}%
\right) ,  \label{lsmix2}
\end{equation}%
where%
\begin{equation}
\Omega =\left( 
\begin{array}{cc}
M_{\sigma _{B}}^{2} & \varepsilon \\ 
\varepsilon & M_{f_{B}}^{2}%
\end{array}%
\right) =\left( 
\begin{array}{cc}
2\alpha +\frac{1}{3}\gamma _{S}-\frac{2\sqrt{2}}{3}z_{S} & -\frac{1}{3}z_{S}+%
\frac{\sqrt{2}}{3}\gamma _{S} \\ 
-\frac{1}{3}z_{S}+\frac{\sqrt{2}}{3}\gamma _{S} & 2\beta +\frac{2}{3}\gamma
_{S}+\frac{2\sqrt{2}}{3}z_{S}%
\end{array}%
\right)
\end{equation}%
(note that $M_{\sigma _{B}}^{2}$ and $M_{f_{B}}^{2}$ receive
extra-contributions from (\ref{lsmix}) and therefore modified from Eq. (\ref%
{scalmass})).

The physical states $\sigma $ and $f_{0}\equiv f_{0}(980)$ are then given by%
\begin{equation}
\left( 
\begin{array}{c}
\sigma \\ 
f_{0}(980)%
\end{array}%
\right) =\left( 
\begin{array}{cc}
\cos (\theta _{S}) & \sin (\theta _{S}) \\ 
-\sin (\theta _{S}) & \cos (\theta _{S})%
\end{array}%
\right) \left( 
\begin{array}{c}
\sigma _{B} \\ 
f_{B}%
\end{array}%
\right)  \label{ts}
\end{equation}%
where%
\begin{equation}
B=\left( 
\begin{array}{cc}
\cos (\theta _{S}) & \sin (\theta _{S}) \\ 
-\sin (\theta _{S}) & \cos (\theta _{S})%
\end{array}%
\right) ,\text{ }B\Omega B^{t}=diag\{M_{\sigma }^{2},M_{f_{0}}^{2}\}.
\label{b}
\end{equation}

The experimental fact that $M_{f_{0}}\simeq M_{a_{0}}$ suggests a small
scalar mixing angle $\theta _{S}$. Unfortunately, to determine $\theta _{S}$
from the masses can be misleading because of the large widths of $k$ and $%
\sigma .$ Furthermore, small variation of the $a_{0}$ and $f_{0}$ masses and
mixing can lead to different scenarios. We therefore prefer to determine the
mixing angle form the decay amplitudes (see section 4).

Nevertheless some interesting considerations can be done; exploiting the
relation $\left\langle \Omega \right\rangle =M_{\sigma }^{2}+ M_{f_{0}}^{2}$
(from Eq. (\ref{b}); the matrix $B$ is orthogonal) one finds $\gamma
_{S}=M_{f_{0}}^{2}+M_{\sigma }^{2}-2\cdot M_{k}^{2}$. The sign of $\gamma
_{S}$ strongly depends on the values of $M_{\sigma }$ and $M_{k}.$ One has
then to tune the parameter $z_{S}$ in order to generate the experimental
value $M_{f_{0}}=M_{a_{0}}\simeq 0.98$ GeV, then the mixing angle $\theta
_{S}$ is also fixed\footnote{%
The parameter $z_{S}$ can be found from $Det[\Omega ]=M_{\sigma }^{2}\cdot
M_{f_{0}}^{2}$ (see Eq. (\ref{b})). Indeed, this is a second order equation
in $z_{S,}$ so strictly speaking two solutions are possible. The solution
corresponding to the smallest absolute value $\left\vert \theta
_{S}\right\vert $ is the one that we take into account. The disregarded
solution typically induces a very large mixing, in disagreement with the
phenomenology (see Section 4).}. We then distinguish among 2 possibilities:

a) $\gamma _{S}>0\rightarrow z_{S}<0$ and $\theta _{S}<0;$ in the upper
state $f_{0}$ the bare components $\sigma _{B}$ and $f_{B}$ are in phase,
while in the lower state $\sigma $ they are out of phase.

b) $\gamma _{S}<0\rightarrow z_{S}>0$ and $\theta _{S}>0;$ the phases are
reversed.

The sign of $\theta _{S}$ is important because of destructive or
constructive interference phenomena in the decays of $\sigma $ and $f_{0}.$
However, as stressed above, it cannot be determined from the knowledge of
the masses because of the large uncertainties on their values. We consider
two examples to elucidate this point. Fixing $M_{k}=0.8$ GeV, we have $%
\gamma _{S}>0$ (and $\theta _{S}<0$) for $M_{\sigma }>0.56$ GeV; vice versa
for $M_{\sigma }<0.56$ one has $\gamma _{S}<0$ (and $\theta _{S}>0$). Small
changes of $M_{k}$ generate very different result; fixing $M_{k}=0.72$ (the
pole of E791 \cite{aitala}) one finds $\gamma _{S}>0$ (and $\theta _{S}<0$)
for $M_{\sigma }>0.277$ GeV ( therefore, if $M_{k}\sim 0.7$ GeV the option $%
\theta _{S}<0$ seems favoured). We notice that the present results differ
from those of \cite{maiani}.

In Section 4 we will determine the values of the scalar mixing angle $\theta
_{S}$ from the phenomenology and discuss the implications for the masses.

\subsection{Two-pseudoscalar decay amplitudes}

In the following we report the results for the two-pseudoscalar decay rates.
We use the following notation: for a given decay mode $S\rightarrow
P_{1}P_{2}$ (where $S$ refers to a scalar state and $P_{1}(P_{2})$ to a
pseudoscalar state) the decay width (for kinematically allowed decays) is
written as%
\begin{equation}
\Gamma _{S\rightarrow P_{1}P_{2}}=\frac{p_{S\rightarrow P_{1}P_{2}}}{8\pi
M_{S}^{2}}g_{S\rightarrow P_{1}P_{2}}^{2},  \label{gsp1p2}
\end{equation}%
where $p_{S\rightarrow P_{1}P_{2}}$ is the three-momentum of (one of) the
outgoing particle(s) (in the rest frame of $S$):%
\begin{equation}
p_{S\rightarrow P_{1}P_{2}}=\frac{1}{2M_{S}}\sqrt{%
M_{S}^{4}+(M_{P_{1}}^{2}-M_{P_{2}}^{2})^{2}-2(M_{P_{1}}^{2}+M_{P_{2}}^{2})M_{S}^{2}%
}.
\end{equation}%
Notice that $g_{S\rightarrow P_{1}P_{2}}^{2}$ already includes the charge
multiplicities and the symmetry factors (see Appendix B).

In general, but especially when $M_{S}<M_{P_{1}}+M_{P_{2}},$ the expression (%
\ref{gsp1p2}) has to be modified taking into account the finite width of the
resonance, thus integrating over the mass of the resonance by employing a
suitable mass distribution (like a Breit-Wigner one). However, the
expressions for the coupling constant are left invariant by these
operations. We refer to Appendix B for a brief recall on the connection
between Eq. (\ref{gsp1p2}) and the Lagrangian (\ref{lag}).

In Table 1 we list the coupling constants for the isovector $a_{0}(980)$ and
the isodoublet(s) $k$ to two pseudoscalar mesons as function of the dominant
and subdominant decay strengths $c_{1}$ and $c_{2}$. The expressions in $%
\{...\}$ represent the invariant amplitudes, eventually multiplied by a
factor $2$ for decays into identical particles, while the coefficients in
front of the parenthesis account for charge multiplicities and symmetry
factors (see Appendix B).

In Table 2 we report the results for the bare states $\sigma _{B}$ and $%
f_{B} $; although not physical when $\theta _{S}\neq 0,$ the expressions are
easier to read and allow to understand the role of the subleading decay
mechanism of Fig. 1.b. It is important to note that the decay mode $%
f_{B}\rightarrow \overline{K}K$ is significantly enhanced with respect to $%
a_{0}\rightarrow \overline{K}K$ when $c_{2}/c_{1}>0.$ This fact shows that a
non-negligible and positive ratio $c_{2}/c_{1}$ can explain why the $%
f_{0}\rightarrow \overline{K}K$ coupling is larger than the $%
a_{0}\rightarrow \overline{K}K$ one. This point will be discussed in the
next Section.

When considering the mixed physical states $\sigma $ and $f_{0}(980)$
defined in Eq. (\ref{ts}) the coupling constants are modified as follows:%
\begin{eqnarray}
g_{\sigma \rightarrow P_{1}P_{2}} &=&\left\{ g_{\sigma _{B}\rightarrow
P_{1}P_{2}}\cdot \cos (\theta _{S})+g_{f_{B}\rightarrow P_{1}P_{2}}\cdot
\sin (\theta _{S})\right\}  \notag \\
g_{f_{0}\rightarrow P_{1}P_{2}} &=&\left\{ -g_{\sigma _{B}\rightarrow
P_{1}P_{2}}\cdot \sin (\theta _{S})+g_{f_{B}\rightarrow P_{1}P_{2}}\cdot
\cos (\theta _{S})\right\}  \label{sf}
\end{eqnarray}

Because of their relevance, we report in Table 3 the explicit expressions
for the decays of $\sigma $ and $f_{0}(980)$ into $\pi \pi $ and $\overline{K%
}K$ as derived from Table 2 and Eqs. (\ref{sf}).

\bigskip

\bigskip

\bigskip

\begin{center}
\textbf{Table 1.} Decay coupling constants for $a_{0}$ and $k$

\vspace*{0.5cm} 
\begin{tabular}{|c|c|}
\hline\hline
\ $S\rightarrow P_{1}P_{2}$ & $g_{S\rightarrow P_{1}P_{2}}$ \\ \hline\hline
$a_{0}\rightarrow \overline{K}K$ & $\sqrt{2}\cdot \left\{ \sqrt{2}c_{1}+%
\frac{1}{\sqrt{2}}c_{2}\right\} $ \\ \hline
$a_{0}\rightarrow \pi \eta $ & $\sqrt{1}\cdot \left\{ \frac{2}{\sqrt{3}}c_{1}%
\left[ \sqrt{2}\cos (\theta _{P})+\sin (\theta _{P})\right] +\sqrt{\frac{2}{3%
}}c_{2}\left[ \cos (\theta _{P})-\sqrt{2}\sin (\theta _{P})\right] \right\} $
\\ \hline
$a_{0}\rightarrow \pi \eta ^{\prime }$ & $\sqrt{1}\cdot \left\{ -\frac{2}{%
\sqrt{3}}c_{1}\left[ \cos (\theta _{P})-\sqrt{2}\sin (\theta _{P})\right] +%
\sqrt{\frac{2}{3}}c_{2}\left[ \sqrt{2}\cos (\theta _{P})+\sin (\theta _{P})%
\right] \right\} $ \\ \hline
$k\rightarrow \pi K$ & $\sqrt{3}\cdot \left\{ \sqrt{2}c_{1}+\frac{1}{\sqrt{2}%
}c_{2}\right\} $ \\ \hline
$k\rightarrow K\eta $ & $\sqrt{1}\cdot \left\{ -\sqrt{\frac{2}{3}}c_{1}\left[
\cos (\theta _{P})-\sqrt{2}\sin (\theta _{P})\right] -\sqrt{\frac{1}{6}}c_{2}%
\left[ \cos (\theta _{P})+2\sqrt{2}\sin (\theta _{P})\right] \right\} $ \\ 
\hline
$k\rightarrow K\eta ^{\prime }$ & $\sqrt{1}\cdot \left\{ -\sqrt{\frac{2}{3}}%
c_{1}\left[ \sqrt{2}\cos (\theta _{P})+\sin (\theta _{P})\right] +\sqrt{%
\frac{1}{6}}c_{2}\left[ 2\sqrt{2}\cos (\theta _{P})-\sin (\theta _{P})\right]
\right\} $ \\ \hline
\end{tabular}

\bigskip

\bigskip

\textbf{Table 2.} Decay coupling constants of the bare states $\sigma _{B}$
and $f_{B}$

\vspace*{0.5cm} 
\begin{tabular}{|c|c|}
\hline\hline
\ $S\rightarrow P_{1}P_{2}$ & $g_{S\rightarrow P_{1}P_{2}}$ \\ \hline\hline
$\sigma _{B}\rightarrow \pi \pi $ & $\sqrt{\frac{3}{2}}\cdot \left\{
2c_{1}+2c_{2}\right\} $ \\ \hline
$\sigma _{B}\rightarrow \overline{K}K$ & $\sqrt{2}\cdot \left\{
c_{2}\right\} $ \\ \hline
$\sigma _{B}\rightarrow \eta \eta $ & $\sqrt{\frac{1}{2}}\cdot \left\{ \frac{%
2}{3}(-c_{1}+c_{2})\left( \cos (\theta _{P})-\sqrt{2}\sin (\theta
_{P})\right) ^{2}\right\} $ \\ \hline
$\sigma _{B}\rightarrow \eta ^{\prime }\eta ^{\prime }$ & $\sqrt{\frac{1}{2}}%
\cdot \left\{ \frac{2}{3}(-c_{1}+c_{2})\left( \sqrt{2}\cos (\theta
_{P})+\sin (\theta _{P})\right) ^{2}\right\} $ \\ \hline
$\sigma _{B}\rightarrow \eta \eta ^{\prime }$ & $\sqrt{1}\cdot \left\{ \frac{%
1}{3}(-c_{1}+c_{2})\left( 2\sqrt{2}\cos (2\theta _{P})-\sin (2\theta
_{P})\right) \right\} $ \\ \hline
$f_{B}\rightarrow \pi \pi $ & $\sqrt{\frac{3}{2}}\cdot \left\{ \sqrt{2}%
c_{2}\right\} $ \\ \hline
$f_{B}\rightarrow \overline{K}K$ & $\sqrt{2}\cdot \left\{ \left( \sqrt{2}%
c_{1}+\frac{3}{\sqrt{2}}c_{2}\right) \right\} $ \\ \hline
$f_{B}\rightarrow \eta \eta $ & $\sqrt{\frac{1}{2}}\cdot \left\{ \frac{1}{6}%
\left( 9\sqrt{2}c_{2}+\sqrt{2}(8c_{1}+c_{2})\cos (2\theta
_{P})+4(-c_{1}+c_{2})\sin (2\theta _{P})\right) \right\} $ \\ \hline
$f_{B}\rightarrow \eta ^{\prime }\eta ^{\prime }$ & $\sqrt{\frac{1}{2}}\cdot
\left\{ \frac{1}{6}\left( 9\sqrt{2}c_{2}-\sqrt{2}(8c_{1}+c_{2})\cos (2\theta
_{P})+4(c_{1}-c_{2})\sin (2\theta _{P})\right) \right\} $ \\ \hline
$f_{B}\rightarrow \eta \eta ^{\prime }$ & $\sqrt{1}\cdot \left\{ \frac{1}{6}%
\left( 4(c_{1}-c_{2}\right) \cos (2\theta _{P})+\sqrt{2}(8c_{1}+c_{2})\sin
(2\theta _{P})\right\} $ \\ \hline
\end{tabular}
\end{center}

\bigskip

\bigskip

\bigskip

\begin{center}
\textbf{Table 3.} Decay coupling constants of the physical states $\sigma $
and $f_{0}$

\bigskip

$\vspace*{0.5cm}%
\begin{tabular}{|c|c|}
\hline\hline
\ $S\rightarrow P_{1}P_{2}$ & $g_{S\rightarrow P_{1}P_{2}}$ \\ \hline\hline
$\sigma \rightarrow \pi \pi $ & $\sqrt{\frac{3}{2}}\cdot \left\{ 2c_{1}\cos
(\theta _{S})+c_{2}(2\cos (\theta _{S})+\sqrt{2}\sin (\theta _{S}))\right\} $
\\ \hline
$\sigma \rightarrow \overline{K}K$ & $\sqrt{2}\cdot \left\{ \sqrt{2}%
c_{1}\sin (\theta _{S})+c_{2}(\cos (\theta _{S})+\frac{3}{\sqrt{2}}\sin
(\theta _{S}))\right\} $ \\ \hline
$f_{0}\rightarrow \pi \pi $ & $\sqrt{\frac{3}{2}}\cdot \left\{ -2c_{1}\sin
(\theta _{S})+c_{2}(-2\sin (\theta _{S})+\sqrt{2}\cos (\theta _{S}))\right\} 
$ \\ \hline
$f_{0}\rightarrow \overline{K}K$ & $\sqrt{2}\cdot \left\{ \sqrt{2}c_{1}\cos
(\theta _{S})+c_{2}(-\sin (\theta _{S})+\frac{3}{\sqrt{2}}\cos (\theta
_{S}))\right\} $ \\ \hline
\end{tabular}%
$

\bigskip
\end{center}

\section{Strong decays: numerical results}

The data about the decay widths reported in \cite{pdg} are still not
complete. The resonance $\sigma $ is very broad ($M_{\sigma }=0.4$-$1.2$
GeV, $\Gamma =0.6$-1 GeV), $k$ is still not yet listed (in the recent result
of \cite{buggexp05} the pole is found at $0.76$ GeV, the width is very
large). The masses for the states $a_{0}(980)$ and $f_{0}(980)$ are well
known, $M_{a_{0}}=948.7\pm 1.2$ MeV $M_{f_{0}}=980\pm 10$ MeV, but the
widths are not, between $40$ and $100$ MeV for both states. The presence of
near-threshold $\overline{K}K$ decays complicates the experimental and
theoretical analysis (see discussion in \cite{olleroset,flatte} and below).

Information about (some of) the coupling constants $g_{S\rightarrow
P_{1}P_{2}}^{2}$ has been extracted directly from experiment \cite%
{buggexp05,bugg,buggexp06,flatte}. In the case of $a_{0}(980),$ for
instance, the quantities $g_{a_{0}\rightarrow \pi \eta }^{2},$ $%
g_{a_{0}\rightarrow \overline{K}K}^{2}$ and $M_{a_{0}}$ are free parameters
of the meson-meson scattering amplitudes (generally the Flatt\`{e}
distribution is used \cite{flatte,flatteorig}). They can be determined by
fitting the theoretical amplitudes to the experimental ones. However, as
explained in \cite{flatte}, relatively large differences in the absolute
values of the coupling constants $g_{a_{0}\rightarrow \pi \eta }^{2}$ and $%
g_{a_{0}\rightarrow \overline{K}K}^{2}$ are found in the literature.
Fortunately, the ratio $g_{a_{0}\rightarrow \pi \eta
}^{2}/g_{a_{0}\rightarrow \overline{K}K}^{2}$ shows a stable behavior.

Furthermore, the presence of strong coupling constant for the near-threshold
decay mode $g_{a_{0}\rightarrow \overline{K}K}$ affects also the channel $%
a_{0}\rightarrow \pi \eta $ rendering the experimental width narrower than
the theoretical result obtained using Eq. (\ref{gsp1p2}), see the discussion
in \cite{olleroset}.

In the following we will compare our results with the analysis of
experimental results of Refs. \cite{buggexp05,bugg,buggexp06}, where the
ratios of coupling constants for various decay channels are deduced; it
should be stressed that the results of \cite{buggexp05,bugg,buggexp06} are
not yet conclusive and depend on the choice of the form factors and other
assumptions, for which we refer to the above cited works for a careful
description.

The ratios of coupling constants for the resonances $a_{0}(980)$ and $%
f_{0}(980)$ as reported in \cite{buggexp05,bugg,buggexp06} are: 
\begin{equation}
\text{ }\frac{g_{f_{0}\rightarrow \overline{K}K}^{2}}{g_{f_{0}\rightarrow
\pi \pi }^{2}}=4.21\pm 0.46,\text{ }\frac{g_{f_{0}\rightarrow \overline{K}%
K}^{2}}{g_{a_{0}\rightarrow \overline{K}K}^{2}}=2.15\pm 0.40,\frac{%
g_{a_{0}\rightarrow \pi \eta }^{2}}{g_{a_{0}\rightarrow \overline{K}K}^{2}}%
=0.75\pm 0.11.  \label{expa0f0}
\end{equation}%
These three results are indeed (at least qualitatively) common to (almost)
all the analyses (see \cite{flatte} and Refs. therein, and the recent
experimental analysis of \cite{ablikim2,kloelast}). The largest uncertainty
is about the crossed ratio $g_{f_{0}\rightarrow \overline{K}K}^{2}/$ $%
g_{a_{0}\rightarrow \overline{K}K}^{2}$; however, the enhanced coupling of
the $f_{0}\rightarrow \overline{K}K$ mode with respect to $a_{0}\rightarrow 
\overline{K}K$ is also a stable result. As we will see, all three ratios in (%
\ref{expa0f0}) are all compatible with a sizable $c_{2}/c_{1}.$

Before considering a nonzero value for $c_{2}/c_{1},$ we analyze as a first
step the case $c_{2}=0$, corresponding to the original work of \cite{jaffe}
and described in the recent paper of \cite{bugg}. We fit the only free
parameter $\theta _{S}$ to the first and the second branching ratios of Eq. (%
\ref{expa0f0}); the ratio $g_{a_{0}\rightarrow \pi \eta
}^{2}/g_{a_{0}\rightarrow \overline{K}K}^{2}=0.50$ does not depend on $%
\theta _{S}$, hence not included in the fit (in \cite{bugg} it is $0.4$
because of a slightly different pseudoscalar mixing angle; here it is $%
\theta _{P}=$ $-9.95^{\circ },$ while in \cite{bugg} the value $\theta _{P}=$
$-17.29^{\circ }$ is used).

We find the following value for $\theta _{S}$ (we refer to this case as
Solution A):%
\begin{equation}
\text{Sol. A: }(c_{2}=0),\text{ }\theta _{S}=\pm 21.6^{\circ }\text{; }(%
\frac{\chi ^{2}}{2}=5.17)  \label{solzero}
\end{equation}%
corresponding to the values%
\begin{equation}
\text{ }\frac{g_{f_{0}\rightarrow \overline{K}K}^{2}}{g_{f_{0}\rightarrow
\pi \pi }^{2}}=4.26,\text{ }\frac{g_{f_{0}\rightarrow \overline{K}K}^{2}}{%
g_{a_{0}\rightarrow \overline{K}K}^{2}}=0.86,\text{ }\frac{%
g_{a_{0}\rightarrow \pi \eta }^{2}}{g_{a_{0}\rightarrow \overline{K}K}^{2}}%
=0.50.  \label{solzerovalues}
\end{equation}%
The large and unsatisfactory $\chi ^{2}$ is generated by the mismatch
between experiment and theory for the ratio $g_{f_{0}\rightarrow \overline{K}%
K}^{2}/g_{a_{0}\rightarrow \overline{K}K}^{2}.$ Two scalar mixing angles are
reported in Eq. (\ref{solzero}) because the fitted quantities are symmetric
for $\theta _{S}\rightarrow -\theta _{S}$ when $c_{2}=0$. Our theoretical
ratio for $g_{f_{0}\rightarrow \overline{K}K}^{2}/g_{f_{0}\rightarrow \pi
\pi }^{2} $ as function of $\theta _{S}$ is deduced from Table 3 and reads
for $c_{2}=0 $: 
\begin{equation}
\frac{g_{f_{0}\rightarrow \overline{K}K}^{2}}{g_{f_{0}\rightarrow \pi \pi
}^{2}}=\frac{2}{3}\cot ^{2}(\theta _{S}).
\end{equation}%
We notice that in \cite{bugg} the theoretical ratio is $g_{f_{0}\rightarrow 
\overline{K}K}^{2}/g_{f_{0}\rightarrow \pi \pi }^{2}=$ $\frac{1}{3}\cot
^{2}(\theta _{S})$; for this reason the mixing angle found in \cite{bugg} is 
$\theta _{S}=\pm 15.9^{\circ }$ and differs from our result. The
extra-factor $1/2$ present in the result of \cite{bugg} could not be
verified when evaluating the traces of Eq. (\ref{lag}).

We now consider a non-zero $c_{2}$: in principle one could determine the
ratio $c_{2}/c_{1}$ from the quantity $g_{a_{0}\rightarrow \pi \eta
}^{2}/g_{a_{0}\rightarrow \overline{K}K}^{2}$ alone. In fact, the
theoretical expression depends only on $c_{2}/c_{1}.$ When $c_{2}=0$ the
result as derived from Table 1 is $0.5$, then $g_{a_{0}\rightarrow \pi \eta
}^{2}/g_{a_{0}\rightarrow \overline{K}K}^{2}$ increases very slowly for
increasing $c_{2}/c_{1}.$ The value $g_{a_{0}\rightarrow \pi \eta
}^{2}/g_{a_{0}\rightarrow \overline{K}K}^{2}=0.6$ is reached when $%
c_{2}/c_{1}=0.6,$ but the value $g_{a_{0}\rightarrow \pi \eta
}^{2}/g_{a_{0}\rightarrow \overline{K}K}^{2}=0.75$ corresponds to $%
c_{2}/c_{1}=2.35.$ Because of this strong sensibility it is not practicable
to determine $c_{2}/c_{1}$ in this way. However, a sizable $c_{2}$ improves
the agreement with the experiment.

The $\chi ^{2}$-method is then used to find the free quantities $c_{2}/c_{1}$
and $\theta _{S}$ which correspond to the best description of (\ref{expa0f0}%
) (Solution B):%
\begin{equation}
\text{Sol. B: }c_{2}/c_{1}=0.62,\text{ }\theta _{S}=-12.8^{\circ }\text{; }(%
\frac{\chi ^{2}}{3}=0.65<1).  \label{soluno}
\end{equation}%
The theoretical ratios evaluated with the parameters of Sol. B are:%
\begin{equation}
\frac{g_{f_{0}\rightarrow \overline{K}K}^{2}}{g_{f_{0}\rightarrow \pi \pi
}^{2}}=4.21,\text{ }\frac{g_{f_{0}\rightarrow \overline{K}K}^{2}}{%
g_{a_{0}\rightarrow \overline{K}K}^{2}}=2.28,\text{ }\frac{%
g_{a_{0}\rightarrow \pi \eta }^{2}}{g_{a_{0}\rightarrow \overline{K}K}^{2}}%
=0.60,\text{ }
\end{equation}%
thus in good agreement with the results of (\ref{expa0f0}). The inclusion of
the subdominant decay diagram of Fig. 1.b leads to a clear improvement of
all three ratios without adding $\overline{K}K$ contributions to the wave
functions of the resonances. Notice that within this Solution a negative
value for the scalar mixing angle is preferred.

In \cite{bugg,buggexp06} a large ratio $g_{\sigma \rightarrow \overline{K}%
K}^{2}/g_{\sigma \rightarrow \pi \pi }^{2}=0.6\pm 0.1$ is deduced from
analysis of $\phi \rightarrow \gamma \pi ^{0}\pi ^{0}$ (such a large ratio
can explain a problem in reproducing the overall normalizations; other
explanations are however possible \cite{oller,roca}). With the parameters of
Sol. B (\ref{soluno}) one finds a very small ratio: $g_{\sigma \rightarrow 
\overline{K}K}^{2}/g_{\sigma \rightarrow \pi \pi }^{2}=4.8\cdot 10^{-7}.$
Such a small value is caused by the destructive interference between the $%
\sigma _{B}$ and $f_{B}$ components (see Table 3). If the value of \cite%
{buggexp06} should be confirmed by future experimental analyses, our Sol. B
should be rejected.

We make a second fit by adding to the data in (\ref{expa0f0}) the value $%
g_{\sigma \rightarrow \overline{K}K}^{2}/g_{\sigma \rightarrow \pi \pi
}^{2}=0.6\pm 0.1.$ The minimum of $\chi ^{2}$ is found for (Solution C):%
\begin{equation}
\text{Sol. C: }c_{2}/c_{1}=0.89,\text{ }\theta _{S}=35.8^{\circ }\text{; }(%
\frac{\chi ^{2}}{4}=2.04).  \label{soldue}
\end{equation}

The results derived from the Solutions A, B and C are summarized in Table 4,
where also other coupling ratios are presented. In Sol. C a large branching
ratio $\overline{K}K/\pi \pi $ for the $\sigma $ resonance is found ($0.65$%
), however $g_{f_{0}\rightarrow \overline{K}K}^{2}/g_{a_{0}\rightarrow 
\overline{K}K}^{2}$ gets worse ($1.12$, generating a large $\chi ^{2};$ it
is however still larger than 1).

\bigskip

\begin{center}
\textbf{Table 4.} Comparison of coupling ratios with the analysis of
experimental results of Refs. \cite{bugg,buggexp06} \footnote{%
The results for Sol. A are equal for the two mixing angles $\theta _{S}=\pm
21.6^{\circ },$ with the exception of two ratios: $g_{f_{0}\rightarrow \eta
\eta }^{2}/g_{f_{0}\rightarrow \pi \pi }^{2}$ and $g_{\sigma \rightarrow
\eta \eta }^{2}/g_{\sigma \rightarrow \pi \pi }^{2},$ for which the first
result corresponds to $\theta _{S}=-21.6^{\circ },$ while the secon (in
parenthesys) to $\theta _{S}=21.6^{\circ }.$}

\bigskip

\begin{tabular}{|c|c|c|c|c|}
\hline\hline
Ratios of $g^{2}$ & Sol. A (\ref{solzero}) & Sol. B (\ref{soluno}) & Sol. C (%
\ref{soldue}) & Analysis of Refs. \cite{bugg,buggexp06} \\ \hline\hline
$g_{a_{0}\rightarrow \pi \eta }^{2}/g_{a_{0}\rightarrow \overline{K}K}^{2}$
& $0.50$ & $0.60$ & $0.63$ & $0.75\pm 0.11$ \\ \hline
$g_{f_{0}\rightarrow \overline{K}K}^{2}/g_{f_{0}\rightarrow \pi \pi }^{2}$ & 
$4.26$ & $4.21$ & $4.35$ & $4.21\pm 0.16$ \\ \hline
$g_{f_{0}\rightarrow \overline{K}K}^{2}/g_{a_{0}\rightarrow \overline{K}%
K}^{2}$ & $0.86$ & $2.28$ & $1.11$ & $2.15\pm 0.4$ \\ \hline
$g_{a_{0}\rightarrow \pi \eta ^{\prime }}^{2}/g_{a_{0}\rightarrow \pi \eta
}^{2}$ & $1.01$ & $0.16$ & $0.05$ & - \\ \hline
$g_{f_{0}\rightarrow \eta \eta }^{2}/g_{f_{0}\rightarrow \pi \pi }^{2}$ & $%
1.78$ $(1.57)$ & $1.35$ & $2.30$ & $<0.33$ \\ \hline
$g_{\sigma \rightarrow \overline{K}K}^{2}/g_{\sigma \rightarrow \pi \pi
}^{2} $ & $0.10$ & $4.8\cdot 10^{-7}$ & $0.65$ & $0.6\pm 0.1$ \\ \hline
$g_{\sigma \rightarrow \eta \eta }^{2}/g_{\sigma \rightarrow \pi \pi }^{2}$
& $0.30$ $(0.02)$ & $0.05$ & $0.10$ & $0.20\pm 0.04$ \\ \hline
$g_{k\rightarrow \pi K}^{2}/g_{\sigma \rightarrow \pi \pi }^{2}$ & $1.16$ & $%
0.78$ & $0.58$ & $(2.14\pm 0.28)$ to $(1.35\pm 0.10)$ \\ \hline
$g_{k\rightarrow \eta K}^{2}/g_{k\rightarrow \pi K}^{2}$ & $0.17$ & $0.12$ & 
$0.11$ & $0.06\pm 0.02$ \\ \hline
$g_{k\rightarrow \eta ^{\prime }K}^{2}/g_{k\rightarrow \pi K}^{2}$ & $0.16$
& $0.006$ & $\sim 0$ & $0.29\pm 0.29$ \\ \hline
\end{tabular}
\end{center}

\bigskip

Some comments are in order:

$a)$ The first three ratios refer to $a_{0}(980)$ and to $f_{0}(980),$ which
have been studied in details in the literature for what concern the $%
\overline{K}K,$ $\pi \eta $ and $\pi \pi $ channels. For these reasons our
preferred solution is Sol. B (\ref{soluno}), which is in good agreement with
these three experimental ratios.

The other ratios refer to the broad $\sigma $ and $k$ states or to channels
far from threshold for $a_{0}$ and $f_{0}$, therefore are not free of
ambiguities.

$b)$ The ratio $g_{\sigma \rightarrow \overline{K}K}^{2}/g_{\sigma
\rightarrow \pi \pi }^{2}$ is the main difference between Solutions B and C,
which also generates different values for the scalar angle, negative in the
first case and positive in the second. We notice also a large discrepancy of
all three Solutions for the branching ratio $g_{f_{0}\rightarrow \eta \eta
}^{2}/g_{f_{0}\rightarrow \pi \pi }^{2},$ whose experimental value reported
in \cite{bugg} is smaller than the theoretical results for Jaffe's
tetraquark states. Future checks on these two quantities may help to
disentangle the nature of the light scalar mesons.

$c)$ In Solutions $B$ and $C$ the ratio $c_{2}/c_{1}$ is not small. The
inclusion of the decay mechanism of Fig. 1.b (which can occur with one gluon
as intermediate state) represents an improvement for the phenomenology of
Jaffe's tetraquark states. Indeed, this result has been anticipated in \cite%
{maiani}, where the quantity $c_{2}/c_{1}$ varies between $0.7$ and $1$ (for
comparison, the parameters $a$ and $b$ in \cite{maiani} are $2c_{1}$ and $%
-c_{2}$ in the present work).

In particular, one can notice that the first three entries of Table 4 are
improved in Sol. B when compared to Sol. A (corresponding to $c_{2}=0$),
thus a non-zero and sizable ratio $c_{2}/c_{1}$ is favoured within the here
presented phenomenological analysis.

In Sol. B the ratio $c_{2}/c_{1}=0.62$ lies in between $1/N_{c}=1/3$ and $1.$
Although larger than $1/3,$ it is still clearly smaller than $1$. Sol. B is
still in agreement with expectations from large $N_{c}$ considerations. In
Sol. C one has $c_{2}/c_{1}=0.89,$ which is of order $1.$ Such a value would
imply a substantial violation from large $N_{c};$ this fact constitutes a
further hint in favor of Sol. B.

$d)$ The coupling $g_{k\rightarrow \pi K}^{2}$ is smaller then $g_{\sigma
\rightarrow \pi \pi }^{2}$ in solution $B$ and $C$ ; this is not in accord
with the experimental result describing such a wide $k$ resonance
(curiously, Solution A is better in this respect). Two experimental values
are reported in Table 4 (see \cite{bugg}), the second one does not represent
such a large mismatch with the theoretical values. It is noticeable that a
similar problem exists when describing the scalar sector above $1$ GeV (\cite%
{giacosachiral} and Refs. therein): the experimental large width of the
state $K_{0}(1430)$ cannot be explained theoretically.

$e)$ The knowledge of the mixing angle does not allow us to deduce $%
M_{\sigma }$ and $M_{k}$. However, we can consider reasonable values for $%
M_{\sigma }$ and determine $M_{k}$ for which the mixing angles of Sol. B and
C are realized: if we take $M_{\sigma }=0.45$ GeV, then a scalar mixing
angle $\theta _{S}=-12.8^{\circ }$ (Sol. B (\ref{soluno})) corresponds to $%
M_{k}=0.69$ GeV, while a mixing angle $\theta _{S}=35.8^{\circ }$ (Sol. C (%
\ref{soldue})) to $M_{k}=0.944$ GeV. Similarly, if $M_{\sigma }=0.55$ GeV
the corresponding couples of values ($\theta _{S}=-12.8^{\circ },M_{k}=0.735$
GeV) and $(\theta _{S}=31.8^{\circ },M_{k}=0.953$ GeV$)$ are found. The
present values for the $k$ pole favor a light mass between $0.7$ and $0.8$
GeV \cite{aitala,buggexp05}, therefore Sol. B is in better agreement with
the data. (However, the analysis of \cite{highk} points to a slightly
heavier kaonic state: $M_{k}=841\pm 30_{-73}^{+81},$ therefore caution is
still needed when driving conclusions).

$f)$ Keeping in mind the remarks at the beginning of this section, we
nevertheless discuss some full widths as (naively) calculated from Eq. (\ref%
{gsp1p2}); in \cite{flatte} various values for $g_{a_{0}\rightarrow \pi \eta
}^{2}$ are reported ($=\overline{g}_{\eta }\cdot 8\pi M_{a_{0}}^{2}$ in \cite%
{flatte}). The results vary between $5$ and $10$ GeV$^{2}.$ We take for
simplicity a value in between: $g_{a_{0}\rightarrow \pi \eta }^{2}=7.5$ GeV$%
^{2}.$ For Sol. B such a value corresponds to $c_{1}=1.32$ GeV, resulting in
(values in MeV) $\Gamma _{f_{0}\rightarrow \pi \pi }=136$, $\Gamma
_{a_{0}\rightarrow \pi \eta }=98$, $\Gamma _{\sigma \rightarrow \pi \pi
}=795 $ and $\Gamma _{k\rightarrow K\pi }=251.$ For Sol. C we have $%
c_{1}=1.19$ GeV, implying $\Gamma _{f_{0}\rightarrow \pi \pi }=58$, $\Gamma
_{a_{0}\rightarrow \pi \eta }=98$, $\Gamma _{\sigma \rightarrow \pi \pi
}=1013$ and $\Gamma _{k\rightarrow K\pi }=241.$ Clearly the widths change
substantially when varying $g_{a_{0}\rightarrow \pi \eta }^{2}$ in the
mentioned range. The qualitative picture emerging is the presence in both
cases of two broad states $\sigma $ and $k$ with $\Gamma _{\sigma
\rightarrow \pi \pi }>$ $\Gamma _{k\rightarrow K\pi }$ (but the latter not
so broad as desired), and two narrower partial decay widths for $f_{0}$ and $%
a_{0}$; notice that the ordering $\Gamma _{f_{0}\rightarrow \pi \pi }>\Gamma
_{a_{0}\rightarrow \pi \eta }$ in Sol. B is reversed in Sol. C. The
inclusion of a mass distribution for the calculation of the decay widths,
thus also evaluating the $\overline{K}K$ modes for $a_{0}$ and $f_{0},$ is
planned as a future step but beyond the goal of the present work.

$g)$ Before moving to the two-photon decays, we wish to remind what has been
considered and what omitted in the performed study on strong decays. The two
interaction terms of the Lagrangian (\ref{lag}) are $SU_{V}(3)$-invariant
and correspond to the diagrams of Fig. 1; other flavour symmetric terms,
listed in Eq. (\ref{all}), and direct $SU_{V}(3)$-breaking interactions have
been omitted. They are supposed to represent corrections to the here
presented scenario(s) (large $N_{c}$-suppressed terms and flavor symmetry
breaking corrections) eventually useful for a more quantitative study. By
performing a fit leading to Sol. B we used as experimental quantities the
values reported in \cite{buggexp05,bugg,buggexp06}; as already noticed
above, these three results, although qualitatively similar to other works on
light scalars, depend on particular assumption of form factors. Furthermore,
the errors are only statistical and not systematic. Even more caution is
needed with the other values reported in the right column of Table 4. Being
aware of these limitations of both theoretical and experimental origin, we
however intended to focus on a particular and interesting aspect of
phenomenology of four-quark states encoded in the two diagrams of Fig. 1.

\section{Two-photon decays}

We now turn the attention to the $\gamma \gamma $-decays of the Jaffe's
tetraquark states. As for the strong decays we consider two analogous
channels, where one has photons instead of mesons as final states in Fig 1.
The Lagrangian reads%
\begin{equation}
\mathcal{L}_{em}=c_{1}^{\gamma \gamma }\mathcal{S}_{ij}\left\langle
A^{i}QA^{j}Q\right\rangle F_{\mu \nu }^{2}-c_{2}^{\gamma \gamma }\mathcal{S}%
_{ij}\left\langle A^{i}A^{j}Q^{2}\right\rangle F_{\mu \nu }^{2},
\label{laggg}
\end{equation}%
where $Q=e\cdot diag\{2/3,-1/3,-1/3\}$ is the charge matrix ($e=\sqrt{4\pi
\alpha }$ is the electron charge, $\alpha \simeq 1/137$) and $F_{\mu \nu
}=\partial _{\mu }A_{\nu }-\partial _{\nu }A_{\mu }$ the electromagnetic
field tensor. Note that the convention for the relative sign of the leading
and subleading terms is the same of Eq. (\ref{lag}) (see also Appendix A).

The decay width into two photons reads%
\begin{equation}
\Gamma _{S\rightarrow \gamma \gamma }=16\pi \alpha ^{2}M_{S}^{3}\cdot
g_{S\rightarrow \gamma \gamma }^{2},  \label{sgg}
\end{equation}%
where the nonzero contributions obviously correspond to $S=a_{0}^{0},$ $%
\sigma ,$ $f_{0}.$ The coupling constants for $a_{0}^{0}$ and for the bare
states $\sigma _{B}$ and $f_{B}$ are deduced from (\ref{laggg}) and read:%
\begin{equation}
g_{a_{0}^{0}\rightarrow \gamma \gamma }=\frac{2c_{1}^{\gamma \gamma
}+c_{2}^{\gamma \gamma }}{3\sqrt{2}},\text{ }g_{\sigma _{B}\rightarrow
\gamma \gamma }=\frac{4c_{1}^{\gamma \gamma }+5c_{2}^{\gamma \gamma }}{9},%
\text{ }g_{f_{B}\rightarrow \gamma \gamma }=\frac{2c_{1}^{\gamma \gamma
}+7c_{2}^{\gamma \gamma }}{9\sqrt{2}}
\end{equation}%
When considering the mixed physical states $\sigma $ and $f_{0}(980)$
defined in Eq. (\ref{ts}) the coupling constants are modified as (see also
Eq. (\ref{sf})):%
\begin{eqnarray}
g_{\sigma \rightarrow \gamma \gamma } &=&g_{\sigma _{B}\rightarrow \gamma
\gamma }\cdot \cos (\theta _{S})+g_{f_{B}\rightarrow \gamma \gamma }\cdot
\sin (\theta _{S}),  \notag \\
g_{f_{0}\rightarrow \gamma \gamma } &=&-g_{\sigma _{B}\rightarrow \gamma
\gamma }\cdot \sin (\theta _{S})+g_{f_{B}\rightarrow \gamma \gamma }\cdot
\cos (\theta _{S}).
\end{eqnarray}

The experimental results for the decay width of $a_{0}$ and $f_{0}$ are
given by \cite{pdg}%
\begin{equation}
\text{ }\Gamma _{f_{0}\rightarrow \gamma \gamma }=0.39_{-0.13}^{+0.10}\text{
KeV, }\Gamma _{a_{0}^{0}\rightarrow \gamma \gamma }=0.30\pm 0.10\text{ KeV.}
\label{ggkev}
\end{equation}%
The experimental value for $\Gamma _{a_{0}^{0}\rightarrow 2\gamma }$ is not
reported as an average in \cite{pdg}; however, it is in accord with the
quoted averages $\Gamma _{a_{0}^{0}\gamma 2\gamma }\cdot (\Gamma _{\eta \pi
}/\Gamma _{tot})=0.24_{-0.07}^{+0.08}$ KeV and $\Gamma _{\overline{K}%
K}/\Gamma _{\eta \pi }=0.183\pm 0.024.$ Combining these two results one
finds $\Gamma _{a_{0}^{0}\rightarrow \gamma \gamma }=0.28\pm 0.10$ KeV, well
compatible with (\ref{ggkev}). In \cite{achasov} it was shown that
two-photon widths about $0.3$ keV are compatible with a tetraquark nature of
the states.

Expressing the results of (\ref{ggkev}) in terms of squared coupling
constants we find%
\begin{equation}
\frac{\Gamma _{a_{0}^{0}\rightarrow \gamma \gamma }}{\Gamma
_{f_{0}\rightarrow \gamma \gamma }}=\frac{g_{a_{0}^{0}\rightarrow \gamma
\gamma }^{2}}{g_{f_{0}\rightarrow \gamma \gamma }^{2}}=0.77\pm 0.48
\label{gqgg}
\end{equation}%
(where $M_{a_{0}}=M_{f_{0}}=0.98$ GeV and an average error of $0.115$ KeV
for $\Gamma _{f_{0}\rightarrow \gamma \gamma }$ have been used).

Large $N_{c}$ results are still valid by invoking vector meson dominance,
for which the decay into two photons occurs via two virtual vector mesons.
Then, we first discuss the case $c_{2}^{\gamma \gamma }=0$ (Solution A),
where only the dominant contribution is considered. By using the mixing
angle $\theta _{S}=21.6^{\circ }$ (see Eq. (\ref{solzero})) we find the
totally wrong ratio $g_{a_{0}^{0}\rightarrow \gamma \gamma
}^{2}/g_{f_{0}\rightarrow \gamma \gamma }^{2}=724.7,$ while the other option 
$\theta _{S}=-21.6^{\circ }$ implies $g_{a_{0}^{0}\rightarrow \gamma \gamma
}^{2}/g_{f_{0}\rightarrow \gamma \gamma }^{2}=2.3$, which is still
unacceptable when compared to the experimental result of (\ref{gqgg}). Even
by modifying the scalar angle $\theta _{S}$ the situation is not improved;
one finds a minimum for the ratio $g_{a_{0}^{0}\rightarrow \gamma \gamma
}^{2}/g_{f_{0}\rightarrow \gamma \gamma }^{2}=1$ at $\theta
_{S}=-70.5^{\circ }.$ Such a value for the scalar mixing angle is however
ruled out by the phenomenology of the strong decay analyzed in the previous
section (a small mixing angle $\theta _{S}$ is common to all three analyzed
scenarios and is in accord with the mass degeneracy of $a_{0}(980)$ and $%
f_{0}(980)$). Therefore, also the two-photon decay shows that the dominant
decay mechanism, corresponding to a switch of $q$ and $\overline{q}$
analogous to Fig. 1.a, is not enough to describe the experimental data.

Let us then consider the case of a non-zero $c_{2}^{\gamma \gamma }.$ We
keep the scalar mixing angle $\theta _{S}$ as found in Sol. B and C
respectively and we determine the ratio $c_{2}^{\gamma \gamma
}/c_{1}^{\gamma \gamma }$ in order to reproduce the experimental result of
Eq. (\ref{gqgg})\footnote{%
Once the ratio is determined, one can fix the strength of $c_{1}^{\gamma
\gamma }$ to reproduce $\Gamma _{f_{0}\rightarrow \gamma \gamma }$ (Eq. (\ref%
{ggkev})). The other width $\Gamma _{f_{0}\rightarrow \gamma \gamma }$ is
then also correctly described.} 
\begin{equation}
\text{Sol. B (Eq. (\ref{soluno}), }\theta _{S}=-12.8^{\circ }\text{)}%
\Rightarrow \frac{c_{2}^{\gamma \gamma }}{c_{1}^{\gamma \gamma }}=0.73
\label{solguno}
\end{equation}%
\newline
\begin{equation}
\text{Sol. C (Eq. (\ref{soldue}), }\theta _{S}=35.8^{\circ }\text{)}%
\Rightarrow \frac{c_{2}^{\gamma \gamma }}{c_{1}^{\gamma \gamma }}=-1.04
\label{solggdue}
\end{equation}

In both cases a large ratio $c_{2}^{\gamma \gamma }/c_{1}^{\gamma \gamma }$
is found; while in Sol. B it is still safely smaller than unity, in Sol. C a
large (and negative $<-1$) ratio is obtained.

Furthermore, in Solution B we find a noticeable correspondence between the
electromagnetic and strong transitions: $c_{2}^{\gamma \gamma
}/c_{1}^{\gamma \gamma }=0.73\sim c_{2}/c_{1}=0.62.$ The two-photon decay,
which in the framework of vector meson dominance is mediated by two virtual
vector mesons, differs from the two-pseudoscalar decay only in the final
stage, therefore the contribution of the subleading diagram with respect to
the leading one is naively expected to be of the same magnitude (and with
the same sign). On the contrary in Solution C we have $c_{2}^{\gamma \gamma
}/c_{1}^{\gamma \gamma }=-1.04<<$ $c_{2}/c_{1}=0.89:$ this fact would mean
very different contributions of the subleading diagram in the decay
amplitude into two vector mesons (then converting into two photons) from the
corresponding transition into two pseudoscalar mesons. The composite
approach used in this work does not allow for a more microscopic insights to
deal rigorously with this issue, therefore this discussion does not
represent a proof in favour of Sol. B. We simply limit to notice that within
Sol. B one has $c_{2}^{\gamma \gamma }/c_{1}^{\gamma \gamma }\sim
c_{2}/c_{1},$ which on the contrary does not take place in Sol. C.

The experimental situation concerning $\sigma \rightarrow 2\gamma $ is less
clear; no average or fit is presented in \cite{pdg}, however two experiments
listed in \cite{pdg} find large $\gamma \gamma $ decay widths: $3.8\pm 1.5$
keV and $5.4\pm 2.3$ keV, respectively. In a footnote it is then state that
these values could be assigned to $f_{0}(1370)$ (actually, in a older
version of Particle Data Group \cite{pdg2000} these values were assigned to
the resonance $f_{0}(1370)$)$.$ It is not clear if the $\gamma \gamma $
signal comes from the high mass tail of the broad $\sigma $ state or from $%
f_{0}(1370)$ (or even from both; in such case the experimental result would
represent the sum, i.e. an upper limit for both resonances). Furthermore,
the application of Eq. (\ref{sgg}) is too naive for the broad $\sigma $
state; in fact, due to the third power of the mass of the decaying
resonance, the two-photon decay width is strongly influenced by a large
width, especially from the right mass tail (the precise form of a mass
distribution would be necessary for a more precise and quantitative
analysis). Here we simply report the results for the ratio $%
g_{a_{0}^{0}\rightarrow \gamma \gamma }^{2}/g_{\sigma \rightarrow \gamma
\gamma }^{2}$ in the two scenarios: $g_{a_{0}^{0}\rightarrow \gamma \gamma
}^{2}/g_{\sigma \rightarrow \gamma \gamma }^{2}$ is $0.83$ in Solution B (%
\ref{solguno}) and $0.42$ in Solution C (\ref{solggdue}). However, if a
two-photon width between $3$ and $5$ keV should be entirely assigned to the $%
\sigma $-resonance as recently discussed in Ref. \cite{pennington}, the
interpretation of the $\sigma $ as a tetraquark state would be problematic,
because in the four-quark scenario the $\gamma \gamma $ decay width of the $%
\sigma $ would be of the same order of the $a_{0}^{0}(980)$ one ($\sim 0.3$
keV) , as the above reported coupling ratios $g_{a_{0}^{0}\rightarrow \gamma
\gamma }^{2}/g_{\sigma \rightarrow \gamma \gamma }^{2}$ suggest. Future work
on the $\sigma \rightarrow \gamma \gamma $ transition is needed both
theoretically and experimentally in order to clarify this crucial issue
about light scalar mesons.

\section{Summary and Conclusions}

The Jaffe's four-quark states have appealing characteristics to be good
candidates for the description of the scalar nonet below $1$ GeV; this paper
aimed to analyze this possibility by studying strong and the electromagnetic
decays of scalar resonances.

We first considered the strong decays of the light scalar mesons below 1
GeV. Beyond the OZI-superallowed decay, in which the scalar tetraquark state
falls apart into two pseudoscalar mesons as depicted in Fig. 1.a, the next
to leading order in the $N_{c}$ expansion, which occurs with (only) one
intermediate gluon as shown in Fig. 1.b, has been systematically taken into
account for all decay coupling constants and presented in Tables 1, 2 and 3.
The two decay mechanisms are described by corresponding terms in an
effective composite interaction Lagrangian and are parametrized by two
strength parameters $c_{1}$ and $c_{2}$ respectively.

As a first step we studied the case $c_{2}=0$, in which the diagram of Fig.
1.b is switched off. The large ratio $g_{f_{0}\rightarrow \overline{K}%
K}^{2}/g_{a_{0}\rightarrow \overline{K}K}^{2}=2.15\pm 0.40$ $>1$ reported in
the experimental analysis of \cite{buggexp06} cannot be described because
the theoretical ratio is $\leq 1$ for $c_{2}=0$. We referred to the case $%
c_{2}=0$ as `Solution A'.

We then allowed for a non-zero ratio $c_{2}/c_{1}$: by fitting $c_{2}/c_{1}$
and the scalar-isoscalar mixing angle $\theta _{S}$ to the three values
reported in Eq. (\ref{expa0f0}) deduced in the analysis of \cite{buggexp06}
and involving the $\overline{K}K$, $\pi \eta $ and $\pi \pi $ channels for
the $a_{0}(980)$ and $f_{0}(980)$ states, a sizable ratio $c_{2}/c_{1}=0.62$
is obtained. The subdominant transition of Fig. 1.b is non negligible in the
present analysis and improves the theoretical description of the mentioned
decay modes of $a_{0}(980)$ and $f_{0}(980)$. The mixing angle $\theta
_{S}=-12.8^{\circ }$ is small and negative: the bare components $\sigma
_{B}\equiv $\textquotedblleft $\overline{u}u\overline{d}d$%
\textquotedblright\ and $f_{B}\equiv $\textquotedblleft $\overline{s}s(%
\overline{u}u-\overline{d}d)$\textquotedblright\ are out of phase in the
physical state $\sigma $ and in phase in the resonance $f_{0}(980).$ We
called this scenario `Solution B', which represents our preferred solution,
as discussed in Sections 3 and 4.

Within Solution B one finds a very small ratio $g_{\sigma \rightarrow
KK}^{2}/g_{\sigma \rightarrow \pi \pi }^{2}\sim 10^{-7},$ while the result
reported in the analysis of Ref. \cite{bugg} reads $0.6\pm 0.1.$ Solution C
has been then built by including this ratio when minimizing the $\chi $%
-squared. As a result, agreement with \cite{buggexp06} is obtained for the
branching ratio $\overline{K}K/\pi \pi $, but the ratio $g_{f_{0}\rightarrow 
\overline{K}K}^{2}/g_{a_{0}\rightarrow \overline{K}K}^{2}$ is found to be $%
1.11$ within Solution C, which is, although larger than 1, clearly worsened
with respect to Solution B. Future experimental check on $g_{\sigma
\rightarrow KK}^{2}/g_{\sigma \rightarrow \pi \pi }^{2}$ may help to clarify
this point. The mixing angle in Sol. C is positive: $\theta _{S}=31.8^{\circ
},$ i.e. the bare components are in phase for $\sigma $ and out of phase for 
$f_{0}(980).$ The sign difference of $\theta _{S}$ is responsible of such
different results concerning the quantity $g_{\sigma \rightarrow
KK}^{2}/g_{\sigma \rightarrow \pi \pi }^{2}.$

The results of Solutions A, B and C are summarized in Table 4, where also
other ratios of squared coupling constants are listed.

We then considered the two-photon decay of the scalar resonances. The
diagrams for such transitions are similar to those of Fig. 1.a and 1.b but
with photons instead of mesons as final states. The corresponding
intensities are parametrized by $c_{1}^{\gamma \gamma }$ and $c_{2}^{\gamma
\gamma }$ respectively.

The discussions follows the same line as before: the case $c_{2}^{\gamma
\gamma }=0$ (Sol. A) is analyzed first. The experimental result $\Gamma
_{a_{0}^{0}\rightarrow 2\gamma }/\Gamma _{f_{0}\rightarrow 2\gamma
}=g_{a_{0}^{0}\rightarrow \gamma \gamma }^{2}/g_{f_{0}\rightarrow \gamma
\gamma }^{2}=0.77\pm 0.48$ cannot be obtained if $c_{2}^{\gamma \gamma }=0$,
thus again pointing to a non-zero contribution of the subleading diagram
with one intermediate gluon.

We then analyzed Solutions B and C: by taking the scalar mixing angle $%
\theta _{S}$ from Solution B or C the ratio $c_{2}^{\gamma \gamma
}/c_{1}^{\gamma \gamma }$ has been determined in order to reproduce the
experimental value $g_{a_{0}^{0}\rightarrow \gamma \gamma
}^{2}/g_{f_{0}\rightarrow \gamma \gamma }^{2}=0.77\pm 0.48.$ In case B one
has the noticeable correspondence $c_{2}^{\gamma \gamma }/c_{1}^{\gamma
\gamma }\sim c_{2}/c_{1},$ that is the contribution of the subleading decay
is similar (and with the same sign) in strong and electromagnetic
transitions. This fact does not take place in Solution C where the two
ratios are large but with opposite sign.

The interpretation of the light scalar states as tetraquark objects is a
viable and interesting possibility. The inclusion of the subleading decay
mechanism of Fig. 1.b improves the description of experimental data for both
strong and electromagnetic transitions within our Solution B.

Future work is however needed: on a experimental side a clear experimental
establishment of the resonance $k$ and a better understanding of the
parameters describing the broad states $\sigma $ and $k$ would constitute a
decisive improvement.

We would like to mention possible future theoretical studies: the decays of
heavier states, such as tensor mesons or scalar mesons above 1 GeV, into two
light scalar mesons as $\sigma \sigma ,$ $kk...$ should be evaluated by
explicitly taking into account the four-quark nature of the light scalars.
Such a analysis would therefore extend and complete the ones in \cite%
{giacosachiral,tensor}. In particular, the fact that the tensor meson nonet
is well established and well known experimentally would offer a suitable
environment for such a study. Furthermore, the two-body decay of excited
pseudoscalar states (eventually mixed with a pseudoscalar glueball) would
also constitute an interesting development. In the recent work of \cite%
{kalashnikova} attention is focused on electromagnetic decays of the type $%
S\rightarrow V\gamma ,$ which may also be helpful in disentangling the
nature of the light scalar states below 1 GeV.

\section*{Acknowledgments}

The author thanks J. Schaffner-Bielich for a careful reading of the
manuscript and for very instructive comments. This research was supported by
the Virtual Institute VH-VI-041 "Dense Hadronic Matter \& QCD Phase
Transitions" of the Helmholtz Association.

\appendix

\section{Tetraquark scalar nonet}

The wave function for a "good" diquark is \cite{exotica} 
\begin{equation}
\left\vert qq\right\rangle =\left\vert \text{space: }L=0\right\rangle
\left\vert \text{spin: }S=0\right\rangle \left\vert \text{color: }\overline{3%
}_{C}\right\rangle \left\vert \text{flavor: }\overline{3}_{F}\right\rangle .
\label{diq}
\end{equation}%
The state $\left\vert qq\right\rangle $ is antisymmetric by exchange of the
two particles in accord with the Pauli principle; the parity quantum number
is $P=(-1)^{L}=+1$. In the following we consider the flavor decomposition
for a diquark (and for an antidiquark). To this end we define the vector $q$
such that $q^{t}=(u,d,s)$; the antisymmetric $\overline{3}_{F}$
decomposition of a diquark is described by the antisymmetric matrix $D$:%
\begin{equation}
D=\sqrt{\frac{1}{2}}(q_{1}\cdot q_{2}^{t}-q_{2}\cdot q_{1}^{t})=\varphi
_{i}A^{i}  \label{d}
\end{equation}%
where the matrices $A^{i}$ are the three antisymmetric $3\times 3$ real
matrices of Eq. (\ref{A}) and the quantities $\varphi _{i}$ are given by%
\begin{eqnarray}
\varphi _{1} &=&\sqrt{\frac{1}{2}}\left( u_{1}d_{2}-u_{2}d_{1}\right) \equiv 
\sqrt{\frac{1}{2}}\left[ u,d\right] , \\
\varphi _{2} &=&\sqrt{\frac{1}{2}}\left( u_{1}s_{2}-u_{2}s_{1}\right) \equiv 
\sqrt{\frac{1}{2}}\left[ u,s\right] , \\
\varphi _{3} &=&\sqrt{\frac{1}{2}}\left( d_{1}s_{2}-d_{2}s_{1}\right) \equiv 
\sqrt{\frac{1}{2}}\left[ d,s\right] .
\end{eqnarray}%
The anti-diquark is then described by the matrix 
\begin{equation}
D^{\dag }=\overline{\varphi }_{i}(A^{i})^{\dag }=-\overline{\varphi }%
_{i}A^{i}
\end{equation}
with $\overline{\varphi }_{1}=\sqrt{1/2}\left[ \overline{u},\overline{d}%
\right] ,$ $\overline{\varphi }_{2}=\sqrt{1/2}\left[ \overline{u},\overline{s%
}\right] $ and $\overline{\varphi }_{3}=\sqrt{1/2}\left[ \overline{d},%
\overline{s}\right] .$

The matrix $D$ has the following transformation properties:%
\begin{eqnarray}
SU(3)\text{-flavor}\text{: } &&D\rightarrow UDU^{t},\text{ }U\in SU(3),
\label{su(3)} \\
P\text{-parity}\text{: } &&D\rightarrow D, \\
C\text{-Charge Conjugation}\text{: } &&D\rightarrow D^{\dag };
\end{eqnarray}%
The flavor transformation (\ref{su(3)}), corresponding to $q\rightarrow Uq$,
follows directly from the definition (\ref{d}).

The matrix $\mathcal{S}$ defined in Eq. (\ref{S}) and entering in Eq. (\ref%
{lag}) is given by $\mathcal{S}_{ij}=\varphi _{i}\overline{\varphi }_{j}$.

In a generic Lagrangian term involving linear tetraquark scalar states (i.e.
one matrix $D$ and one matrix $D^{\dag }$) one has a schematic form like 
\begin{equation}
\left( ...D...D^{\dag }...\right) =\mathcal{S}_{ij}\left(
...A^{i}...(A^{j})^{\dag }...\right)  \label{oper}
\end{equation}%
where dots indicate operation with matrices, such as multiplication, traces,
etc. The form on the l.h.s. is useful to check the $SU(3)$-flavor
transformations, while the form in the r.h.s. allows the calculations of the
decay amplitudes (see Tables 1, 2, 3 and Appendix B). When describing the
scalar $\rightarrow $ two-pseudoscalar decay we then have the following
possible flavor-invariant terms:%
\begin{eqnarray}
\mathcal{L}_{S\rightarrow PP} &=&-c_{1}\left\langle D\mathcal{P}^{t}D^{\dag }%
\mathcal{P}\right\rangle +c_{2}\left\langle DD^{\dag }\mathcal{P}%
^{2}\right\rangle +c_{3}\left\langle DD^{\dag }\mathcal{P}\right\rangle
\left\langle \mathcal{P}\right\rangle  \notag \\
&&+c_{4}\left\langle DD^{\dag }\right\rangle \left\langle \mathcal{P}%
^{2}\right\rangle +c_{5}\left\langle DD^{\dag }\right\rangle \left\langle 
\mathcal{P}\right\rangle ^{2},  \label{all}
\end{eqnarray}%
where $\mathcal{P}=\frac{1}{\sqrt{2}}\sum_{i=0}^{8}P^{i}\lambda _{i}.$ The
first two terms corresponds to the two diagrams of Fig. 1, i.e. to the last
two terms of Eq. (\ref{lag}) (by applying the operation (\ref{oper}) the
equality is easily seen) described throughout all the work. The third term
is also suppressed by a factor $N_{c}$ only with respect to the
OZI-superallowed decay, but the transition can occur with at least two
transverse gluons as intermediate states attached to the pseudoscalar
flavour-singlet state. The fourth and the fifth terms are further suppressed.

The Lagrangian $\mathcal{L}_{S\rightarrow PP}$ is clearly charge and parity
invariant (we recall that $\mathcal{P\rightarrow }U\mathcal{P}U^{\dag },$ $%
\mathcal{P\rightarrow -P}$ and $\mathcal{P\rightarrow P}^{t}$ under flavor,
parity and charge conjugation transformations).

Formally the two-photon Lagrangian (\ref{laggg}) can be obtained from Eq. (%
\ref{all}) by replacing $\mathcal{P}\rightarrow QF_{\mu \nu }$, $%
c_{i}\rightarrow c_{i}^{\gamma \gamma }$ (and then contracting the Lorentz
indices). In virtue of $\left\langle Q\right\rangle =0$ the first, the
second and the fourth terms `survive' the replacing; then, neglecting the
latter we get the Lagrangian (\ref{laggg}) (by invoking vector meson
dominance, large $N_{c}$ arguments can be applied also for the two-photon
decays).

When considering the quadratic term for the scalar-tetraquark states care is
needed; let us introduce an extra-index $a=1,2$ with $D^{a}=\varphi
_{i}^{a}A^{i}.$ The elements $\mathcal{S}_{ij}$ (Eq. (\ref{S})) are formed
by the two corresponding diquark $\varphi _{i}^{a}$ and antidiquark $%
\overline{\varphi }_{j}^{a}$: $\mathcal{S}_{ij}=\varphi _{i}^{a}\overline{%
\varphi }_{j}^{a}.$ Let us consider the flavor-invariant term $\left\langle
D^{1}\left( D^{2}\right) ^{\dag }\right\rangle \left\langle D^{2}\left(
D^{1}\right) ^{\dag }\right\rangle ;$ we aim to show that this is nothing
else but $4\left\langle \mathcal{S}^{2}\right\rangle $ (proportional to the
mass term of Eq. (\ref{lag}) in the flavour-symmetric limit):%
\begin{equation*}
\left\langle D^{1}\left( D^{2}\right) ^{\dag }\right\rangle \left\langle
D^{2}\left( D^{1}\right) ^{\dag }\right\rangle =\varphi _{i}^{1}\overline{%
\varphi }_{j}^{2}\varphi _{k}^{2}\overline{\varphi }_{l}^{1}\left\langle
A^{i}A^{j}\right\rangle \left\langle A^{k}A^{l}\right\rangle
\end{equation*}%
\begin{equation}
=4(\varphi _{i}^{1}\overline{\varphi }_{l}^{1})(\varphi _{k}^{2}\overline{%
\varphi }_{j}^{2})\delta _{ij}\delta _{kl}=4\mathcal{S}_{il}\mathcal{S}%
_{li}=4\left\langle \mathcal{S}^{2}\right\rangle .
\end{equation}%
Then, in Eq. (\ref{lag}) modification from the flavor symmetric limit are
parametrized by the matrix $\chi _{S}=diag\{\alpha ,\beta ,\beta \}$ and $%
\alpha <\beta .$

In Section 2.2 (Eq. (\ref{lsmix})) we considered extra-terms in the scalar
singlet-octet and singlet-singlet channels. The singlet-singlet channel,
proportional to $S_{0}^{2}$, is flavour-symmetric and corresponds to $%
\left\langle D^{1}\left( D^{1}\right) ^{\dag }\right\rangle \left\langle
D^{2}\left( D^{2}\right) ^{\dag }\right\rangle .$ The singlet-octet mixing
introduced in Eq. (\ref{lsmix}) breaks flavour invariance. Here we notice
that a flavour symmetric term, also including singlet-octet mixing (but
affecting also the isovector and isodoublet states) can be constructed for
the quadratic scalar sector: 
\begin{equation}
\left\langle D^{1}\left( D^{1}\right) ^{\dag }D^{2}\left( D^{2}\right)
\right\rangle =\mathcal{S}_{il}\mathcal{S}_{kl}\left\langle
A^{i}A^{j}A^{k}A^{l}\right\rangle .
\end{equation}%
Further study on the mass sector for tetraquark scalar states including this
term could be interesting.

\section{Strong coupling constants}

We briefly recall the connection between the Lagrangian (\ref{lag}) and the
coupling constants $g_{S\rightarrow P_{1}P_{2}}^{2}$ entering in Eq. (\ref%
{gsp1p2}). After evaluating the traces, the Lagrangian (\ref{lag}) can be
rewritten as a sum over all the decay channels $S\rightarrow P_{1}P_{2}$,
where $S=\{a_{0}^{\pm },a_{0}^{0},k^{\pm },k_{0},\overline{k}_{0},\sigma
,f_{0}\}$ and $P=\{\pi ^{\pm },\pi ^{0},K^{\pm },K_{0},\overline{K}_{0},\eta
,\eta ^{\prime }\}$:%
\begin{eqnarray}
\mathcal{L}_{S\rightarrow PP} &\mathcal{=}&c_{1}\mathcal{S}_{ij}\left\langle
A^{i}\mathcal{P}^{t}A^{j}\mathcal{P}\right\rangle -c_{2}\mathcal{S}%
_{ij}\left\langle A^{i}A^{j}\mathcal{P}^{2}\right\rangle  \notag \\
&=&\mathcal{(}\sqrt{2}c_{1}+\frac{c_{2}}{\sqrt{2}}\mathcal{)(}%
a_{0}^{0}K^{-}K^{+}-K^{0}\overline{K}^{0})\mathcal{+}(c_{1}+c_{2})(2\sigma
_{B}\pi ^{+}\pi ^{-}+\sigma _{B}(\pi ^{0})^{2})+....  \notag \\
&=&\sum_{S,P_{1},P_{2}}\lambda _{S\rightarrow P_{1}P_{2}}SP_{1}P_{2},
\end{eqnarray}%
where in the second line we wrote the $a_{0}^{0}$ to $\overline{K}K$ and the 
$\sigma _{B}$ to $\pi \pi $ couplings explicitly; they serve as illustrative
examples of the adopted conventions. The coupling constants at this stage
(without sum/average over isomultiplets) are simply given by $%
g_{S\rightarrow P_{1}P_{2}}=s\cdot \lambda _{S\rightarrow P_{1}P_{2}},$
where $s=2/\sqrt{2}=\sqrt{2}$ when $P_{1}=P_{2},$ $1$ otherwise.

We have, for instance, $g_{a_{0}^{0}\rightarrow K^{+}K^{-}}=\mathcal{(}\sqrt{%
2}c_{1}+\sqrt{1/2}c_{2}\mathcal{)},$ $g_{a_{0}^{0}\rightarrow \overline{K}%
^{0}K^{0}}=-\mathcal{(}\sqrt{2}c_{1}+\sqrt{1/2}c_{2}\mathcal{)},$ $g_{\sigma
_{B}\rightarrow \pi ^{+}\pi ^{-}}=$ $2(c_{1}+c_{2}),$ $g_{\sigma
_{B}\rightarrow \pi ^{0}\pi ^{0}}=\sqrt{2}(c_{1}+c_{2}).$ Notice that by
plugging these coupling constants into (\ref{gsp1p2}) we get the partial
decay width for the corresponding channel, such as $\Gamma
_{a_{0}^{0}\rightarrow K^{+}K^{-}}$.

Let us then group the final states in corresponding isomultiplets: $P=\{\pi
,K,\eta ,\eta ^{\prime }\}.$ One has to perform the sum over the members of
the isomultiplets, for instance:%
\begin{eqnarray}
g_{a_{0}^{0}\rightarrow \overline{K}K}^{2} &=&g_{a_{0}^{0}\rightarrow
K^{+}K^{-}}^{2}+g_{a_{0}^{0}\rightarrow \overline{K}^{0}K^{0}}^{2}%
\rightarrow g_{a_{0}^{0}\rightarrow \overline{K}K}=\sqrt{2}%
g_{a_{0}^{0}\rightarrow K^{+}K^{-}} \\
g_{\sigma _{B}\rightarrow \pi \pi }^{2} &=&g_{\sigma _{B}\rightarrow \pi
^{+}\pi ^{-}}^{2}+g_{\sigma _{B}\rightarrow \pi ^{0}\pi ^{0}}^{2}\rightarrow
g_{\sigma _{B}\rightarrow \pi \pi }=\sqrt{\frac{3}{2}}g_{\sigma
_{B}\rightarrow \pi ^{+}\pi ^{-}}
\end{eqnarray}

The sign of the resulting coupling constant can be chosen arbitrarily for
the isotriplet and isodoublets $a_{0}$ and $k$ states. In the $a_{0}^{0}$ to 
$\overline{K}K$ case we chose the sign of the charged decay channel $%
K^{+}K^{-}$ (which is opposite to the $\overline{K}^{0}K^{0}$ one); Table 1
is compiled following this convention. This however does not affect the
decay rates expressed in (\ref{gsp1p2}). Care is needed in the
scalar-isoscalar sector because of mixing occurring in Eq. (\ref{sf}), such
as $g_{\sigma _{B}\rightarrow \pi \pi }:$ the sign of $g_{\sigma
_{B}\rightarrow \pi \pi }$ is taken to be the same as $g_{\sigma
_{B}\rightarrow \pi ^{+}\pi ^{-}}$ and $g_{\sigma _{B}\rightarrow \pi
^{0}\pi ^{0}}.$ In the scalar-isoscalar sector the various contributions to
the sum over final isomultiplets have the same overall sign, thus making the
definition unambiguous. Table 2 and 3 follow this convention.

As a last step an average over the initial isospin multiplets is performed,
that is we consider $S=\{a_{0},k,\sigma ,f_{0}\}.$ For instance, $%
g_{a_{0}\rightarrow \overline{K}K}^{2}$ is given by:%
\begin{equation}
g_{a_{0}\rightarrow \overline{K}K}^{2}=\frac{1}{3}\left(
g_{a_{0}^{0}\rightarrow \overline{K}K}^{2}+g_{a_{0}^{+}\rightarrow \overline{%
K}K}^{2}+g_{a_{0}^{-}\rightarrow \overline{K}K}^{2}\right)
=g_{a_{0}^{0}\rightarrow \overline{K}K}^{2}.
\end{equation}%
The terms of the average are equal (this is true for each decay of $a_{0}$
and $k;$ for $\sigma $ and $f_{0}$ no average is needed).

\bigskip

\bigskip

\newpage

\begin{figure}[tbp]
\begin{center}
\leavevmode
\leavevmode
\vspace{4.3cm} \includegraphics{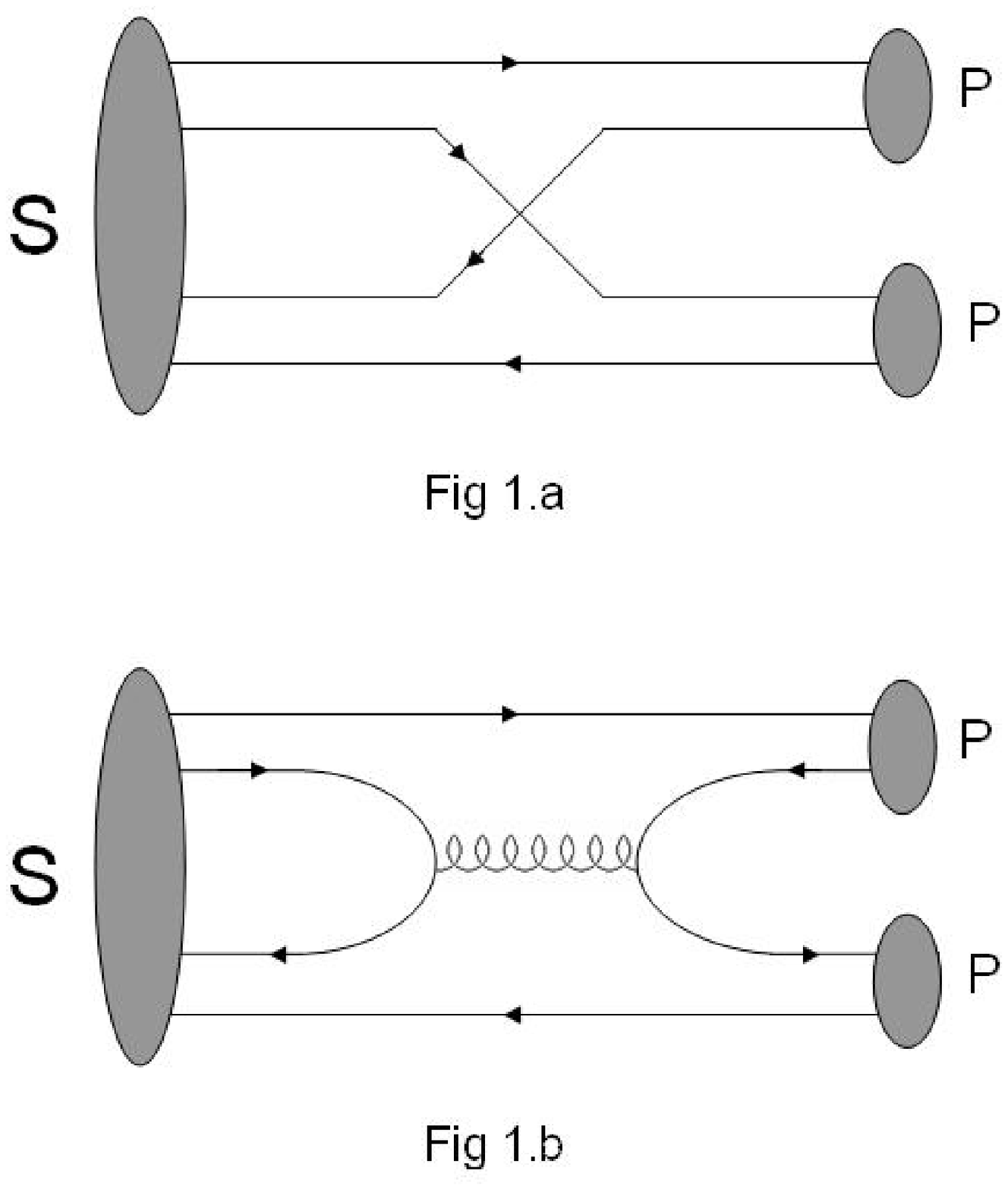}
\end{center}
\par
\vspace{5.4cm}
\caption{{Dominant (1.a) and subdominant (1.b) contributions to the
transition amplitudes of a scalar tetraquark state into two pseudoscalar
mesons.}}
\label{Fig1}
\end{figure}

\bigskip

\bigskip


\bigskip

\begin{thebibliography}{99}
\bibitem{amslerrev} C.~Amsler and N.~A.~Tornqvist, 
Phys.\ Rept.\ \textbf{389}, 61 (2004). 

\bibitem{closerev} F.~E.~Close and N.~A.~Tornqvist, 
J.\ Phys.\ G \textbf{28}, R249 (2002) [arXiv:hep-ph/0204205]; 

\bibitem{mink} P.~Minkowski and W.~Ochs, 
Nucl.\ Phys.\ Proc.\ Suppl.\ \textbf{121}, 123 (2003)
[arXiv:hep-ph/0209225]; 

\bibitem{close95} C.~Amsler and F.~E.~Close, 
Phys.\ Lett.\ B \textbf{353}, 385 (1995) [arXiv:hep-ph/9505219]; 
C.~Amsler and F.~E.~Close, 
Phys.\ Rev.\ D \textbf{53}, 295 (1996) [arXiv:hep-ph/9507326]. 

\bibitem{weing} W.~J.~Lee and D.~Weingarten, 
Phys.\ Rev.\ D \textbf{61}, 014015 (2000) [arXiv:hep-lat/9910008]; 

\bibitem{gutsche} M.~Strohmeier-Presicek, T.~Gutsche, R.~Vinh Mau and
A.~Faessler, 
Phys.\ Rev.\ D \textbf{60}, 054010 (1999) [arXiv:hep-ph/9904461]. 

\bibitem{closekirk} F.~E.~Close and A.~Kirk, 
Eur.\ Phys.\ J.\ C \textbf{21}, 531 (2001) [arXiv:hep-ph/0103173]. 

\bibitem{giacosa} F.~Giacosa, T.~Gutsche and A.~Faessler, 
Phys. Rev. C \textbf{71}, 025202 (2005) [arXiv:hep-ph/0408085]. 

\bibitem{giacosachiral} F.~Giacosa, T.~Gutsche, V.~E.~Lyubovitskij and
A.~Faessler, 
Phys.\ Rev.\ D \textbf{72}, 094006 (2005) [arXiv:hep-ph/0509247]. 

F.~Giacosa, T.~Gutsche, V.~E.~Lyubovitskij and A.~Faessler, 
Phys.\ Lett.\ B \textbf{622}, 277 (2005) [arXiv:hep-ph/0504033]. 

\bibitem{mesonicmol1} M.~B.~Voloshin and L.~B.~Okun, 
JETP Lett.\ \textbf{23}, 333 (1976) [Pisma Zh.\ Eksp.\ Teor.\ Fiz.\ \textbf{%
23}, 369 (1976)]. 
K.~Maltman and N.~Isgur, 
Phys.\ Rev.\ Lett.\ \textbf{50}, 1827 (1983). 
K.~Maltman and N.~Isgur, 
Phys.\ Rev.\ D \textbf{29}, 952 (1984). 

\bibitem{mesonicmol2} N.~A.~Tornqvist, 
Phys.\ Rev.\ Lett.\ \textbf{67}, 556 (1991). 

N.~A.~Tornqvist, 
Z.\ Phys.\ C \textbf{61}, 525 (1994) [arXiv:hep-ph/9310247]. 

T.~E.~O.~Ericson and G.~Karl, 
Phys.\ Lett.\ B \textbf{309}, 426 (1993). 

\bibitem{jaffe} R.~L.~Jaffe, 
Phys.\ Rev.\ D \textbf{15} (1977) 267. 

R.~L.~Jaffe, 
Phys.\ Rev.\ D \textbf{15} (1977) 281. 

R.~L.~Jaffe and F.~E.~Low, 
Phys.\ Rev.\ D \textbf{19}, 2105 (1979). 

\bibitem{exotica} R.~L.~Jaffe, 
Phys.\ Rept.\ \textbf{409} (2005) 1 [Nucl.\ Phys.\ Proc.\ Suppl.\ \textbf{142%
} (2005) 343] [arXiv:hep-ph/0409065]. 

\bibitem{maiani} L.~Maiani, F.~Piccinini, A.~D.~Polosa and V.~Riquer, 
Phys.\ Rev.\ Lett.\ \textbf{93} (2004) 212002 [arXiv:hep-ph/0407017]. 

\bibitem{pdg} S.~Eidelman \textit{et al.} [Particle Data Group
Collaboration], 
Phys.\ Lett.\ B \textbf{592}, 1 (2004). 

\bibitem{vanbeveren} E.~Van Beveren, T.~A.~Rijken, K.~Metzger, C.~Dullemond,
G.~Rupp and J.~E.~Ribeiro, 
Z.\ Phys.\ C \textbf{30} (1986) 615. 

\bibitem{ishida} S.~Ishida, M.~Ishida, T.~Ishida, K.~Takamatsu and T.~Tsuru, 
Prog.\ Theor.\ Phys.\ \textbf{98} (1997) 621 [arXiv:hep-ph/9705437]. 

\bibitem{black} D.~Black, A.~H.~Fariborz, F.~Sannino and J.~Schechter, 
Phys.\ Rev.\ D \textbf{58} (1998) 054012 [arXiv:hep-ph/9804273]. 

\bibitem{aitala} E.~M.~Aitala \textit{et al.} [E791 Collaboration], 
Phys.\ Rev.\ Lett.\ \textbf{86} (2001) 765 [arXiv:hep-ex/0007027]. 

\bibitem{buggexp05} D.~V.~Bugg, 
arXiv:hep-ex/0510014. 

\bibitem{bugg} D.~V.~Bugg, 
arXiv:hep-ph/0603089. 

\bibitem{buggexp06} D.~V.~Bugg, 
arXiv:hep-ex/0603023. 

\bibitem{ablikim} M.~Ablikim \textit{et al.} [BES Collaboration], 
Phys.\ Lett.\ B \textbf{598} (2004) 149 [arXiv:hep-ex/0406038]. 

\bibitem{dobado} A.~Dobado and J.~R.~Pelaez, 
Phys.\ Rev.\ D \textbf{56} (1997) 3057 [arXiv:hep-ph/9604416]. 

\bibitem{olleroset} J.~A.~Oller and E.~Oset, 
Nucl.\ Phys.\ A \textbf{620} (1997) 438 [Erratum-ibid.\ A \textbf{652}
(1999) 407] [arXiv:hep-ph/9702314]. 

\bibitem{ollerosetpelaez} J.~A.~Oller, E.~Oset and J.~R.~Pelaez, 
Phys.\ Rev.\ D \textbf{59} (1999) 074001 [Erratum-ibid.\ D \textbf{60}
(1999) 099906] [arXiv:hep-ph/9804209]. 

\bibitem{pelaezprl} J.~R.~Pelaez, 
Phys.\ Rev.\ Lett.\ \textbf{92} (2004) 102001 [arXiv:hep-ph/0309292]. 

\bibitem{pelaez} J.~R.~Pelaez, 
Mod.\ Phys.\ Lett.\ A \textbf{19} (2004) 2879 [arXiv:hep-ph/0411107]. 

\bibitem{jaffelatt} M.~G.~Alford and R.~L.~Jaffe, 
Nucl.\ Phys.\ B \textbf{578} (2000) 367 [arXiv:hep-lat/0001023]. 

\bibitem{okiharu} F.~Okiharu, H.~Suganuma, T.~T.~Takahashi and T.~Doi, 
arXiv:hep-lat/0601005. 

\bibitem{fariborz} A.~H.~Fariborz, R.~Jora and J.~Schechter, 
Phys.\ Rev.\ D \textbf{72} (2005) 034001 [arXiv:hep-ph/0506170]. 
A.~H.~Fariborz, 
Int.\ J.\ Mod.\ Phys.\ A \textbf{19} (2004) 2095 [arXiv:hep-ph/0302133]. 

\bibitem{napsuciale} M.~Napsuciale and S.~Rodriguez, 
Phys.\ Rev.\ D \textbf{70} (2004) 094043 [arXiv:hep-ph/0407037]. 

\bibitem{chpt} 
S.~Weinberg, 
Physica A \textbf{96} (1979) 327; 
J.~Gasser and H.~Leutwyler, 
Annals Phys.\ \textbf{158}, 142 (1984); 
Nucl.\ Phys.\ B \textbf{250} (1985) 465. 


\bibitem{scherer} S.~Scherer, 
arXiv:hep-ph/0210398. 

\bibitem{ecker} G.~Ecker, J.~Gasser, A.~Pich and E.~de Rafael, 
Nucl.\ Phys.\ B \textbf{321}, 311 (1989); 
G.~Ecker, J.~Gasser, H.~Leutwyler, A.~Pich and E.~de Rafael, 
Phys.\ Lett.\ B \textbf{223}, 425 (1989). 

\bibitem{cirigliano} V.~Cirigliano, G.~Ecker, H.~Neufeld and A.~Pich, 
JHEP \textbf{0306} (2003) 012 [arXiv:hep-ph/0305311]. 

\bibitem{venugopal} E.~P.~Venugopal and B.~R.~Holstein, 
Phys.\ Rev.\ D \textbf{57}, 4397 (1998) [arXiv:hep-ph/9710382]. 

\bibitem{juergen} J.~Schaffner-Bielich and J.~Randrup, 
Phys.\ Rev.\ C \textbf{59} (1999) 3329 [arXiv:nucl-th/9812032]. 

J.~Schaffner-Bielich, 
Phys.\ Rev.\ Lett.\ \textbf{84} (2000) 3261 [arXiv:hep-ph/9906361]. 


\bibitem{klempt} E.~Klempt, 
arXiv:hep-ex/0101031. 

\bibitem{Benayoun:1999au} M.~Benayoun, L.~DelBuono and H.~B.~O'Connell, 
Eur.\ Phys.\ J.\ C \textbf{17} (2000) 593 [arXiv:hep-ph/9905350]. 

\bibitem{bini} C.~Bini, 
eConf \textbf{C030626} (2003) FRAT08 [arXiv:hep-ex/0308046]. 

\bibitem{tensor} F.~Giacosa, T.~Gutsche, V.~E.~Lyubovitskij and A.~Faessler, 
Phys.\ Rev.\ D \textbf{72} (2005) 114021 [arXiv:hep-ph/0511171]. 

\bibitem{flatte} V.~Baru, J.~Haidenbauer, C.~Hanhart, A.~Kudryavtsev and
U.~G.~Meissner, 
Eur.\ Phys.\ J.\ A \textbf{23} (2005) 523 [arXiv:nucl-th/0410099]. 

\bibitem{flatteorig} S.~M.~Flatte, 
Phys.\ Lett.\ B \textbf{63} (1976) 228. 

\bibitem{ablikim2} M.~Ablikim \textit{et al.} [BES Collaboration], 
Phys.\ Lett.\ B \textbf{607} (2005) 243 [arXiv:hep-ex/0411001]. 

\bibitem{highk} M.~Ablikim \textit{et al.} [BES Collaboration], 
Phys.\ Lett.\ B \textbf{633} (2006) 681 [arXiv:hep-ex/0506055]. 

\bibitem{kloelast} F.~Ambrosino \textit{et al.} [KLOE Collaboration], 
Phys.\ Lett.\ B \textbf{634} (2006) 148 [arXiv:hep-ex/0511031]. 

\bibitem{oller} J.~A.~Oller, 
Nucl.\ Phys.\ A \textbf{714} (2003) 161 [arXiv:hep-ph/0205121]. 

\bibitem{roca} J.~E.~Palomar, L.~Roca, E.~Oset and M.~J.~Vicente Vacas, 
Nucl.\ Phys.\ A \textbf{729} (2003) 743 [arXiv:hep-ph/0306249]. 

\bibitem{achasov} N.~N.~Achasov, S.~A.~Devyanin and G.~N.~Shestakov, 
Phys.\ Lett.\ B \textbf{96} (1980) 168. 

\bibitem{pdg2000} D.~E.~Groom \textit{et al.} [Particle Data Group
Collaboration], 
Eur.\ Phys.\ J.\ C \textbf{15}, 1 (2000). 

\bibitem{pennington} M.~R.~Pennington, 
arXiv:hep-ph/0604212. 

\bibitem{kalashnikova} Y.~Kalashnikova, A.~Kudryavtsev, A.~V.~Nefediev,
J.~Haidenbauer and C.~Hanhart, 
Phys.\ Rev.\ C \textbf{73} (2006) 045203 [arXiv:nucl-th/0512028]. 
\end{thebibliography}
\end{document}